\documentclass[useAMS,usenatbib]{mn2e}

\usepackage{psfig}
\usepackage{epsfig}
\usepackage{graphicx}

\title[Be/X-ray binaries and candidates: catalogue]{Be/X-ray binaries and candidates}
\author[N.V. Raguzova and S.B. Popov]{N.V. Raguzova$^{1}$\thanks{E-mail:
raguzova@sai.msu.ru} and S.B. Popov$^{1}$\thanks{E-mail:
polar@sai.msu.ru}\\
$^{1}$Sternberg Astronomical Institute, Universitetski pr. 13, 119992 Moscow, Russia}

\begin{document}

\date{}

\pagerange{\pageref{firstpage}--\pageref{lastpage}} \pubyear{2005}

\maketitle

\label{firstpage}

\begin{abstract}
We present a compilative catalogue of Be/X-ray binaries and
candidates in the Galaxy and in the Large and Small Magellanic Clouds.
This catalogue contains 130 sources and provides information on  names
and spectral types of optical components,  distances, spin
characteristics of neutron stars, and on orbital and X-ray properties of
binary systems.
We give brief comments on each object and provide necessary references to
original data.
\end{abstract}

\begin{keywords}
Catalogs -- X-rays: binaries -- stars: emission-line, Be
\end{keywords}

\section{Introduction}

Be/X-ray binaries form the largest subclass of High Mass X-ray Binaries
(HMXBs). These objects attract interests of specialists in several branches
of astrophysics: stellar astrophysics, accretion theory, close binary
evolution, etc. In recent years
thanks to new X-ray satellites data on these sources greatly increased
(see a recent review and references in \citet{n2005}).
In this paper (which is an extended and updated version of \citet{pr2004})
we present a compilative catalogue of these
sources and provide brief comments on each source in the sample.

\section{The catalogue}
\label{catalogue}

In the tables below we present a compilative catalogue of Be/X-ray
stars. We separate galactic sources, systems in Large Magellanic Cloud (LMC)
and in the Small Magellanic Cloud (SMC).
In each table sources are ordered according to right ascension.
In the first column we give sources names. When possible, the first name
corresponds to notation in \cite{liu2000}.
In the second and third columns
we present spectral type of the massive companion and its magnitude.
In the forth column we give spin period, and in the fifth --- orbital
period. Then we give orbital eccentricity and distance to the source.
In the 8th column we give $L_\mathrm{max}$ -- the maximal observed
luminosity.
In the last column  pulse fractions are given.
For sources in LMC and SMC we do not show distances, also as for systems in
SMC orbital eccentricities are not known we omit this column for them, too.
Some comments and more detailed description of each object
can be found in subsections "Comments to the tables" below (each subsection
refers to galactic, LMC or SMC sources).
References in the tables are given in square brackets after each quantity  
(in few cases we do not follow the data up to
the first determination of a parameter, but give a
reference to some catalogue).

\subsection{Galactic sources: comments to the table~\ref{tableG}}

\begin{table*}
\caption{Be/X-Ray Binaries and candidates in the Galaxy}
\begin{tabular}{|l|c|c|c|c|c|c|c|c|c|}
\hline

Name        & Spec. & $m_V$ & $P_\mathrm{spin}$, s & $P_\mathrm{orb}$, d &  e
 & d, kpc & $L_\mathrm{max}$, & Pulse   \\
 & type    &      &                            &
 &       &              &  erg s$^{-1}$ & frac., \% \\

\hline

 & & & & & & & &  \\

0053+604       & B0.5 IVe [1] & 1.6-3.0 [2] & & 203.59 [3] & 0.26 [3] & 0.188 [4] & $3.9\cdot 10^{34}$ [5] & \\
($\gamma$ Cas) & & & & & & & & \\

0115+634        & B0.2 Ve [6] & 15.5 [6] & 3.6 [5] & 24.3 [6] & 0.34 [6] & 7-8 [6] & $3.0\cdot 10^{37}$ [5] & 40-60 [7]\\

J0146.9+6121  & B1 III-Ve [8] & 11.2 [8] & 1412 [5] & & & 2.5 [5] & $3.5\cdot 10^{35}$ [5] & \\
(V831 Cas)     & & & & & & & & \\

0236+610        & B0.5 Ve [9] & 10.7 [10] & & 26.45 [10] & & 3.1 [5] & $2\cdot 10^{34}$ [5] & \\
(V615 Cas)     & & & & & & & & \\

0331+530        & O8-9 Ve [11] & 15.7 [11] & 4.4 [5] & 34.3 [11] & 0.3 [11] & 7 [1] & $ \ga 10^{38}$ [11] & \\
(BQ Cam)       & & & & & & & & \\

0352+309        & O9.5 IIIe & 6.1-6.8 [12] & 837 [12] & 250 [12] & 0.11 [12] & 1.3 [12] & $3\cdot 10^{35}$ [12] &  \\
(X Per)            & -B0 Ve [12]& & & & & & & \\

J0440.9+4431  & B0 III-Ve [13] & 10.78 [13] & 203 [5] & & & 3.2 [5] & $3\cdot 10^{34}$ [5] & \\

0535+262         & B0 IIIe [14] & 8.9-9.6 [10] & 105 [5] & 111 [10] & 0.47 [1] & 2.4 [5] & $2\cdot 10^{37}$ [5] & 20-100 \\
(V725 Tau)      & & & & & & & &  [15], [16] \\

0556+286 & B5ne [10] & 9.2 [10] & & & & & &  \\

J0635+0533 & B2V-B1IIIe [10] & 12.83 [10] & 0.0338 [10] & & &  $2.5-5$ [17] & $(9$-$35)\cdot 10^{33}$ [17] & $\sim$20 [18] \\

0726-260 & O8-9Ve [10] & 11.6 [10] & 103.2 [10] & 34.5 [10] & & $\sim$6 [19] & $2.8 \cdot 10^{35}$ [19] &  $\sim 30$ [19] \\

0739-529 & B7 IV-Ve [10] & 7.62 [10] & & & & & &   \\

0749-600 & B8 IIIe [10] & 6.73 [10] & & & & & &   \\

J0812.4-3114 &  B0.5 V-IIIe [10] & 12.42 [10] & 31.89 [10] & 80 [20] & & 9 [21] & $1.1\cdot 10^{36}$ [21] &  \\

0834-430 & B0-2 III-Ve [10] & 20.4 [10] & 12.3 [10] & 105.8 [1] & 0.12 [1] & 5 [5] & $1.1\cdot 10^{37}$[5] &  $<$15 [22] \\

J1008-57 & O9e-B1e [1] & 15.27 [10] & 93.5 [10] & 247.5 [1] & 0.66 [1] & 2 [5] & $2.9 \cdot 10^{35}$ [5] & 60 [23] \\

1036-565 & B4 IIIe [10] & 6.64 [10] & & & & & &   \\

J1037.5-5647 & B0V-IIIe [10] & 11.3 [10] & 862 [10] & & & 5 [5] & $4.5\cdot 10^{35}$ [5] &  52 [24] \\

1118-615 & O9.5 V-IIIe [10] & 12.1 [10] & 405 [10] & & & 5 [5] &  $5\cdot 10^{36}$ [5]& \\

1145-619 & B1 Vne [10] & 9.3 [10] & 292.4 [10] & 187.5 [10] & $>$0.5 [1] & 0.5 [5] & $7.4 \cdot 10^{34}$ [5] &  28-70 [25] \\

1249-637 & B0 IIIe [10]& 5.31 [10] & & & & & &   \\

1253-761 & B7 Vne [10] & 6.49 [10] & & & & & &   \\

1255-567 & B5 Ve [10] & 5.17 [10] & & & & & &  \\

1258-613 & B2 Vne [10] & 13.5 [10] & 272 [10] & 132.5 [1] & $>$0.5 [1] &  2.4 [5] & $1\cdot 10^{36}$ [5] &  \\
(GX 304-1) & & & & & & &  &  \\

1417-624 &  B1 Ve [1] & 17.2 [10] & 17.6 [10] & 42.12 [10] & 0.446 [1] & 10 [5] & $8 \cdot 10^{36}$ [5]  &  \\

J1452.8-5949 & & & 437.4 [10] & & & 9 [5] &  $8.7\cdot 10^{33}$ [5] & 50-100 [26]\\

J1543-568 &  & & 27.1 [1] & 75.6 [27] & $<$0.03 [27] & $>10$ [27] & $>10^{37}$ [27] &   60-70 [27] \\

1553-542 & Be? & & 9.26 [1] & 30.6 [10] & $<$0.09 [1] & 10 [5] & $7 \cdot 10^{36}$ [5] & 30 [28] \\

1555-552 & B2nne [10] & 8.6 [10] & & & & & &  \\

J170006-4157 & & & 714.5 [10] & & & 10 [5] &  $7.2 \cdot 10^{34}$ [5] & $\sim 30$ [29] \\

J1739-302 &  & & & & & 8.5 [30] & $4.2\cdot 10^{37}$ [30] &  \\

J1739.4-2942 & Be? & & & & & & &  \\

J1744.7-2713 & B2 V-IIIe [10] & 8.4 [10] & & & & 0.9 [31] & $1.8\cdot 10^{32}$ [31] & \\

J1749.2-2725 & Be? & & 220.38 [10] & & & 8.5 [5,32] &  $2.6 \cdot 10^{35}$ [5,32] &  \\

J1750-27 & & & 4.45 [10] & 29.8 [10] & & & & \\

J1820.5-1434 & O9.5-B0 Ve [33] & & 152.26 [35] & & &  4.7 [5] & $9 \cdot 10^{34}$ [5] &  33 [35]\\

1843+00 & B0-B2 & 20.9 [36] & 29.5 [10] & & & $>10$ [36] & $3\cdot 10^{37}$ [37] & 7 [38]  \\
                 & BIV-Ve [36]  &  & & & & & &   \\

1845-024 & & & 94.8 [10] & 242.18 [1] & 0.88 [10] & 10 [5] & $6\cdot 10^{36}$ [5] &  \\

J1858+034 & & & 221 [10] & & & & & 25 [39] \\

1936+541 & Be [10] & 9.8 [10] & & & & & & \\

J1946+274 & B0-B1  & 18.6 [40] & 15.8 [10] & 169.2 [40] & 0.33 [40] & 5 [5] & $5.4\cdot 10^{36}$ [5] & 30 [41] \\
                      & BIV-Ve [40]  &  & & & & & &   \\

J1948+32 & B0 Ve [42] & 14.2 [43] & 18.76 [1] & 40.4 [42] & 0.03 [42] & 9.5 [42] & $2.1\cdot 10^{37}$ [42] & 55-70 [42] \\

2030+375 & B0e [1] & 19.7 [10] & 41.8 [1] & 46.03 [10] & 0.41 [1] & 5 [5] & $1 \cdot 10^{38}$ [5] &  36 [7]\\

J2030.5+4751 & B0.5 V-IIIe [10] & 9.27 [10] & & & & 2.7 [31] & $1.7\cdot 10^{33}$ [31] &  \\

J2058+42 & Be [47] & & 198 [10] & 110 [10] & & 7 [5] &  $2\cdot 10^{36}$ [5] &  36 [7]\\

2103.5+4545 & B0 Ve [44] & 14.2 [44] & 358.6 [44] & 12.7 [44] & $\sim 0.4$ [44] & 6.5 [44] & $ 3 \cdot 10^{36}$ [44] &  45-80 [48] \\

2138+568 & B1 V-B2 Ve [45] & 14.2 [45] & 66.3 [1] & & & 3.8 [5] &  $9.1 \cdot 10^{35}$ [5] & 5-85 [45]  \\
(Cep X-4) & & & & & & & &   \\

2206+543 & B1e [10] & 9.9 [10] & 392 [10] & 9.57 [1] & & 2.5 [5] & $2.5 \cdot 10^{35}$~[5]& \\

2214+589 & Be [10] & 11 [10] & & & & & &  \\

J2239.3+6116 &  B0 V - B2 IIIe [10] & 15.1 [10] & 1247 [46] & 262.6 [1]  & & 4.4 [10] & $\sim 2.3\cdot 10^{36}$ [46] & 40 [46] \\

\hline
\label{tableG}
\end{tabular}
\end{table*}

References to the table 1:\\
(1) \cite{z2002};
(2) \cite{j03};
(3) \cite{hhs00};
(4) \cite{plk97};
(5) \cite{hs00};
(6) \cite{no01};
(7) \cite{hwf04};
(8) \cite{rnc00};
(9) \cite{hc81};
(10) \cite{liu2000};
(11) \cite{nrf99};
(12) \cite{dlp01};
(13) \cite{yit00};
(14) \cite{nrf00};
(15) \cite{bks98};
(16) \cite{ffm85};
(17) \cite{kph1999};
(18) \cite{cmn2000};
(19) \cite{cp1997};
(20) \cite{rnb2001};
(21) \cite{rr1999a};
(22) \cite{whs1997};
(23) \cite{pg1994};
(24) \cite{rr99};
(25) \cite{wfs1999a};
(26) \cite{oop1999};
(27) \cite{icm2001};
(28) \cite{kra1983};
(29) \cite{torii1999};
(30) \cite{smith1998};
(31) \cite{mhd1997};
(32) \cite{torii1998};
(33) \cite{icp2000}.
(34) \cite{mlt97};
(35) \cite{kth1998};
(36) \cite{inc2001};
(37) \cite{pss2000};
(38) \cite{whs1997};
(39) \cite{tcm1998};
(40) \cite{wfc2003};
(41) \cite{pam2001};
(42) \cite{tl2005};
(43) \cite{nim2003};
(44) \cite{rnf2004};
(45) \cite{wfs1999b};
(46) \cite{isc2001};
(47) \cite{wwf2005};
(48) \cite{fdb2005}

\noindent
{\bf 0053+604.} ($\gamma$ Cas, 3A 0053+604, BD+59 144, HD 5394, LS~I
$+60\degr133$, 2S 0053+604, 1H 0053+604, 4U 0054+60)
$\gamma$~Cassiopeiae is one of the best known Be stars; it was the first
emission-line star discovered by Angelo Secchi in 1866, and it has
spectral classification of B0 IVe. Its visual magnitude varies between about 3.0 and
1.6, although usually it stays around 2.5. This object is one of the ROSAT bright sources and
also was detected by IRAS. $\gamma$ Cas has long been known to be very variable
in  optics  and it is also a moderately strong X-ray source with a
luminosity of the order of $10^{33}\, \mathrm{erg~s^{-1}}$
(\citealt{mws76}; \citealt{wsh82}). Such a luminosity would not be surprising for X-ray
emission from an early type star of spectral type O or B --- some active
early type stars have a similar luminosity (\citealt{cwm94};
\citealt{kmt94}). However, the hardness of the X-ray emission of
$\gamma$~Cas is extraordinarily high in comparison with usual X-ray emission from
early type stars. If we fit the spectrum with a thermal model the
resultant temperature is roughly 10 keV or more (\citealt{hkh94}; \citealt{mki86}).
It is not common for early type stars, and resembles more  spectra of
X-ray pulsars and accreting white dwarf binaries. There are currently two
competing interpretations of the nature of the observed X-ray emission:
one is the accretion of the wind from $\gamma$ Cas onto a white dwarf
companion and the other is that X-rays originate due to some physical processes
in the outer atmosphere of $\gamma$ Cas itself. Arguments for and against these
two hypotheses are best summarized in studies by \cite{kmi98} and
\cite{rs00}.

\noindent
{\bf 0115+634.} (V635 Cas, 1H 0115+635, 4U 0115+63, 3U
0115+63, 2E 0115.1+6328, H 0115+634, 4U 0115+634)
This source is one of the best studied Be/X-ray systems. This transient
was first reported in the Uhuru satellite survey (\citealt{gmg72};
\citealt{fjc78}), though a search in the Vela 5B data base revealed that the source had
already been observed by this satellite since 1969 (\citealt{wrp89}). X-ray
outbursts have been observed from 4U 0115+63
by Uhuru (\citealt{ftj76}), HEAO-1
(\citealt{wdp79}; \citealt{rmh79}), Ginga (\citealt{ttk92}),
CGRO/Batse (\citealt{bcc97}), RXTE
(\citealt{whf99}; \citealt{hc99}; \citealt{crh00}) and reoccured
with intervals from one to several years. Precise positional
determinations by the SAS 3, Ariel V and HEAO-1
satellites (\citealt{ccl78}; \citealt{jbd78})
were used to identify this system with a heavily reddened
Be star with a visual magnitude $V=15.5$
(\citealt{jkc78}; \citealt{hc81b}). \cite{rcc78} used
SAS 3 timing observations to derive the orbital parameters of this binary
system. Due to the fast rotation of the neutron star
centrifugal inhibition of accretion prevents the onset of X-ray emission
unless the ram pressure of accreted material reaches
a relatively high value. Magnetic field of the neutron star
is $1.3\cdot 10^{12}\ \mathrm{G}$ (\citealt{mmn99}).
Pulse fraction was obtained in a model-dependent way in the range 20-50
keV (see \citet{hwf04} for details and references).

\noindent
{\bf J0146.9+6121.} (V831 Cas, 2S 0142+61, RX~J0146.9+6121,  LS I $+61\degr 235$)
RX J0146.9+6121 is an accreting neutron star with a 25 min spin period,
the longest known period of any X-ray pulsar in a Be-star system.
This fact was realized (\citealt{msn93}) only after the re-discovery of
this source in the ROSAT All Sky Survey and its identification with the 11th
magnitude Be star LS I $+61\degr 235$ (\citealt{mbb91}).
Indeed the 25 min periodicity had already been discovered with EXOSAT
(\citealt{wmg87}), but it was attributed to a nearby source 4U 0142+614.
The optical star is probably a member of the open cluster NGC 663 at a
distance of about 2.5 kpc (\citealt{tce91}). For this distance, the 1-20 keV
luminosity during the EXOSAT detection in 1984
was $\sim 10^{36}\, \mathrm{erg~s^{-1}}$ (\citealt{msn93}).
All the observations of RX J0146.9+6121 carried out after its
re-discovery yielded lower luminosities, of the order of a few
$10^{34}\ \mathrm{erg~s^{-1}}$ , until an observation
with the Rossi XTE satellite showed that in July 1997 the flux started
to rise again \citep{ham98}, though not up to the level of the
first EXOSAT observation.

\noindent
{\bf 0236+610.} (V615 Cas, 2E 0236.6+6101, LS I $+61\degr303$, 1E
0236.6+6100, RX J0240.4+6112)
LS I $+61\degr 303$ is a radio emitting X-ray binary which exhibits radio
outbursts every 26.5 d. The radio outburst peak and the outburst phase
are
known to vary over
a time scale of $\sim$ 4 yr (\citealt{gxb89}; \citealt{g99}).  The 26.5~d
period is believed to
be the orbital period. \cite{hc81} confirmed the radio period by analysis
of
three-year observation of radial velocity. They concluded that the
optical
spectrum corresponds
to a rapidly rotating B0 V star. The 4 yr modulation
has been discovered on the basis of continued radio monitoring.

\noindent
{\bf 0331+530.} (BQ Cam, EXO 0331+530, V 0332+53)
EXOSAT observed three outbursts from V0332+53 between 1983
November and 1984 January. Two properties of the system were discovered: 
the 4.4 s spin period and 
a sudden decrease of luminosity at the end of $\sim 1$ month
long recurrent outbursts. 
The latter  was interpreted as an
onset of the centrifugal barrier (\citealt{swd85}; \citealt{swr86}). An
upper limit of $\sim 5 \cdot 10^{33}\, \mathrm{erg~s^{-1}}$  to the
source
quiescent emission (1--15 keV) was derived on that occasion with the EXOSAT
Medium
Energy Detector. Doppler shifts in pulse arrivals indicate that the
pulsar is in orbit around a Be star with a period of 34.3 days and
eccentricity 0.3 (\citealt{swd85}). Observations during a
subsequent outburst with Ginga led to the discovery of a cyclotron
line feature corresponding to  $3\cdot 10^{12}\ \mathrm{G}$ magnetic
field (\citealt{mkk84}).
BeppoSAX and Chandra observations  allowed to study this transient
at the faintest flux levels thus far (\citealt{csi02}).
\cite{csi02} concluded that the quiescent emission of this X-ray
transient likely originates
from accretion onto the magnetospheric boundary of the neutron star in
the propeller regime and/or from deep crustal heating resulting from
pycnonuclear
reactions during the outbursts. Recently, the source was observed by
Integral \citep{kmp2005}. The authors confirm the existence of cyclotron lines:
 the fundamental line at 24.9$\pm$0.1
kev, the first harmonic at 50.5$\pm$0.1 kev as well as the second harmonic
at 71.7$\pm$0.8 kev.

\noindent
{\bf 0352+309.} (X~Per, HD~24534, 3A~0352+309, 2E~0352.2+3054,
H~0352+309, 4U~0352+30,
4U~0352+309, 1H~0352+308, 2A~0352+309, H~0353+30, HD~24534, 3U~0352+30)
The X-ray source 4U~0352+309 is a persistent low luminosity pulsar in a
binary system with the Be star X~Persei (X~Per). Its 837 s pulsation period
was
discovered with the UHURU satellite (\citealt{wms76}; \citealt{wms77}), and
is still one of
the longest periods of any known accreting pulsar (\citealt{bcc97}, and
references therein). X Per
is a bright and highly variable star with a visual magnitude that ranges
from $\sim 6.1$ to
$\sim 6.8$ (\citealt{mbf74}; \citealt{rlt97}). The spectral class has been
estimated
to be O9.5~III to B0~V (\citealt{s82}; \citealt{frc92}; \citealt{lrr97}).
Based on spectroscopic parallax, distance estimates range from $700 \pm
300$ pc up to $1.3 \pm 0.4$ kpc
(\citealt{frc92}; \citealt{lrr97}; \citealt{rlt97}; \citealt{twr98}).
The X-ray luminosity varies on long timescales (years) from
$\sim 3\cdot 10^{35}\ \mathrm{erg~s^{-1}}$ to
$\sim 5\cdot 10^{34}\ \mathrm{erg~s^{-1}}$ (for the assumed distance  1.3
kpc; \citealt{rcf93}). \cite{dlp01} have determined a complete orbital
ephemeris of the system
using data from the Rossi X-ray Timing Explorer (RXTE).
\cite{chg01} have discovered a cyclotron resonant scattering feature at
29 keV in the X-ray spectrum of 4U~0352+309 using observation taken with
the RXTE.
The cyclotron resonant scattering feature energy implies a magnetic
field strength at the polar cap of $3.3\cdot 10^{12}\ \mathrm{G}$.

\noindent
{\bf J0440.9+4431.} (RX J0440.9+4431,  VES 826)
RX J0440.9+4431/BSD 24-491 was confirmed as an accreting
Be/X-ray system following the discovery of X-ray pulsations, with
barycentric pulse
period of $202.5 \pm 0.5$ s from RXTE observations (\citealt{rr99}).

\noindent
{\bf 0535+262.} (V725 Tau, HD 245770, 1A 0535+26, 1H 0536+263, 3A
0535+262, BD+26 883, 4U 0538+26, 1A 0535+262, H 0535+262)
The transient A 0535+26 is one of the best studied Be/X-ray binaries.
This source was discovered in 1975 by Ariel 5 (\citealt{res75}) and
showed a 104~s periodicity  indicating the presence of a highly
magnetized
neutron star.
The optical counterpart was later identified with the Be star HDE~245770
(\citealt{lcj79}) allowing the classification of the source as a Be/X-ray
binary.
The pulsed fraction is 20\% at 30-40 keV and increases significantly with
energy, reaching 100\% at 100 keV (\citealt{ffm85}).
Magnetic field of the neutron star is $4.3\cdot 10^{12}\ \mathrm{G}$
(\citealt{mmn99}).

\noindent
{\bf 0556+286.} (4U 0548+29, 1H 0556+286)
The X-ray source was detected by
HEAO1. Probably earlier it was 
observed by UHURU 4U 0548+29 (\citealt{wood1984}).
No detection was made after that.
A Be-star is known in this direction.

\noindent
{\bf J0635+0533.} (SAX J0635+0533)
Discovered by BeppoSAX (\citealt{kph1999}).
\cite{z2002} gives the spectral classification of the optical counterpart
as B0.5 IIIe.
X-ray luminosity is
$(9-35) \cdot 10^{33}$~erg~s$^{-1}$ (2-10 keV) for $d=2.5-5$~kpc
(\citealt{kph1999}).
Bolometric luminosity (0.1-40 keV) was estimated to be
$1.2 \cdot 10^{35}$~erg~s$^{-1}$ for $d=5$~kpc (\citealt{cmn2000}).
Pulse fraction was obtained by BeppoSAX (2-10 keV).
The source can be identified with the gamma-ray source 2EG J0635+0521.
Low luminosity together with very fast rotation propose that
the neutron star has a low magnetic field (see discussion in \citet{cmn2000}).

\noindent
{\bf 0726-260.} (4U 0728-25, 3A 0726-260, V441 Pup, 1H 0726-259, LS 437)
Detected by many experiments (UHURU, HEAO1, Ariel 5, ROSAT, RXTE).
Pulse fraction was estimated
as $(I_\mathrm{max}-I_\mathrm{min})/(I_\mathrm{max}+I_\mathrm{min})$
from the graph in \cite{cp1997} (RXTE 2-20 keV).
The spectral and photometrical analysis of this source led \cite{nrb1996}
to conclude
that the primary is an O8-9Ve star.

\noindent
{\bf 0739-529.} (1H 0739-529 ) 
Detected by HEAO1 (\citealt{wood1984}).

\noindent
{\bf 0749-600.} (1H 0749-600) Detected by HEAO1 (\citealt{wood1984}).
Situated in the open cluster NGC 2516 (\citealt{liu2000}).

\noindent
{\bf J0812.4-3114.} (RX J0812.4-3114, V572 Pup, LS 992)
RX~J0812.4-3114 was discovered by \cite{mhd1997} during a search
for high-mass X-ray binaries by cross-correlating SIMBAD OB star catalogs
with low Galactic latitude sources from the ROSAT all-sky survey.
Thus, this X-ray
source has an identified optical counterpart, the Be star LS~992, and so
it was suspected that this source belongs to the Be/X-ray binaries.
\cite{rnb2001} classify it as B0.2 IVe.
The X-ray light curve of LS 992/RX~J0812.4-3114
is characterized by 31.88 second
pulsations, while the X-ray spectrum is best
represented by an absorbed power-law
component with a exponentially cut-off (\citealt{rr1999a}).
In December 1997 the source made a transition from
a quiescent state to a flaring state (\citealt{cp2000}),
in which regular flares separated by 80 day intervals
were detected with the All-Sky
Monitor (ASM) on-board the Rossi X-ray Timing Explorer. \cite{cp2000}
attributed
the origin of these flares to the periastron passage of the neutron star,
hence this periodicity was naturally associated with the orbital period.
\cite{cp2000} have found strong evidence for the presence
of a $\sim 80$ day period in the ASM light curve of RX~J0812.4-3114.
By comparison with other Be star X-ray binaries,
the time of maximum flux is likely to coincide with periastron passage of
a neutron star. The orbital period of $\sim 80$ days combined with the $\sim
32$ second pulse period is consistent with
the correlation between orbital and pulse
period that is found for the majority of
Be/neutron star binaries (\citealt{c86}).

\noindent
{\bf 0834-430.} (GS 0834-430)
The hard X-ray transient GS~0834-430 was discovered by the WATCH
experiment on board GRANAT in 1990 at a flux level of about 1 Crab in
the 5-15 keV energy band (see \citealt{whs1997}). The source was later
observed by GINGA (\citealt{mak90a}; \citealt{mak90b})
and ROSAT as a part of the All
Sky Survey (\citealt{hpb90}).
The pulsations at a period of 12.3 s were
observed during the GINGA, ROSAT and ART-P observations
(\citealt{mak90c};
\citealt{ade1992}; \citealt{hpb90}; \citealt{gs91}).
GS 0834-43 was also monitored by BATSE
between April 1991 and July 1998. In particular,
seven outbursts were observed from April 1991
till June 1993 with a peak and intra-outburst
flux of about 300 mCrab and $ < 10$ mCrab,
respectively (\citealt{wfh1997}). The recurrence
time of 105-115 days was interpreted
as the orbital period of the system. However, no further
outbursts have been observed since July 1993
either by CGRO/BATSE or by the the All Sky
Monitor  on board  RXTE. All these findings suggest
that GS 0834-43 is a new Be-star/X-ray binary system with an
eccentric orbit (\citealt{wfh1997}).
Based on both photometric and
spectroscopic findings \cite{icc2000} concluded that optical
counterpart of this X-ray pulsar is most
likely a B0-2 V-IIIe star at a distance of 3-5 kpc.
Pulse fraction was obtained by BATSE (20-50 keV).

\noindent
{\bf J1008-57.} (GRO J1008-57 ) Discovered by BATSE in 1993.
Pulse fraction $\sim$~60\% was
obtained by ROSAT (0.1-2.4 keV) (\citealt{hwf04}).
High-energy data (BATSE: 20-70 keV) gives nearly the same value about
67\% \citep{hwf04}.
Orbital period is uncertain. An estimate of 247.5 days comes from
the best fit of BATSE data (\citealt{no2001}). Other (earlier) estimates
were about 135 days (\citealt{liu2000}).
The counterpart is shown to be an OB star with a strong infrared excess
and Balmer emission lines, suggesting a Be-type primary (\citealt{cre94}).

\noindent
{\bf 1036-565.} (3A 1036-565, 1A 1034-56)
Probably the same object as J1037.5-5647.

\noindent
{\bf J1037.5-5647.} (LS 1698, RX J1037.5-5647)
Discovered by ROSAT in 1997.
Probably the same source as 4U1036-56/3A1036-565.
The source was observed in quiescence
(\citealt{rr99}). $L_\mathrm{min}=1.1\cdot 10^{34}$~erg~s$^{-1}$.
Pulse fraction was obtained by RXTE (3-20 keV).

\noindent
{\bf 1118-615.} (1A 1118-615, 1A 1118-616, WRAY 15-793, 2E 1118.7-6138)
The hard X-ray transient A 1118-615 was discovered serendipitously in
1974
by the Ariel-5 satellite (\citealt{esw75}) during an observation
of Cen~X-3 (4U~1119-603). The same series of observations revealed
pulsations with a period of $405.3 \pm 0.6$ s (\citealt{isb1975}). However,
in
the initial announcement of the discovery of the pulsations, they were
wrongly attributed to an orbital period, suggesting that A~1118-615
consisted
of two compact objects (\citealt{isb1975}).
This hard X-ray transient underwent a major outburst only twice:
in 1974, when it was discovered by Ariel-5 satellite, and from December
1991
to February 1992 (\citealt{bcc97}).
The source was observed by \cite{mjp88} using the Einstein and EXOSAT
observatories in 1979 and 1985 respectively. On both occasions a weak
signal was detected confirming that low-level accretion was occurring.
The correct optical counterpart was identified as the Be star
He~3-640/Wray~793 by \cite{ci75}.
The primary has been classified as O9.5IV-Ve (\citealt{jic81}),
with strong Balmer emission lines
indicating the presence of an extended envelope.
According to \cite{vgp92}, the exact classification is complicated by
many faint absorption and emission lines (mostly of Fe~II), but the
overall
spectrum is found to be similar to that of the optical counterparts to
other
known Be/X-ray sources.
The source was observed by \cite{cp85} at UV wavelengths using the IUE
satellite. They confirmed the identification of the counterpart and
reported
prominent UV lines characteristic of a Be star.
Despite the large observational efforts made during last years and mainly
after the 1991-1992 outburst, the Hen3-640/1A~118-615 system is still
poorly
understood. The orbital period of the system is unknown. Corbet's pulse
period/orbital period diagram (\citealt{c86}) gives an orbital period
estimate of $\sim 350$ days.

\noindent
{\bf 1145-619.} (V801 Cen, 2S 1145-61, 2S 1145-619, 2S 1145-62, LS 2502,
3U 1145-61, 4U 1145-62, 4U 1145-619, 4U 1145-61, 3A 1145-619, 2E 1145.5-6155,
H 1147-62, H 1145-619)
Initially observed by UHURU (together with 1145.1-9141). Two sources were
distinguished by Einstein observatory (HEAO2).
In the paper by \cite{liu2000} the optical counterpart was classified as B1
Vne.
Pulse fraction was obtained by BATSE (20-50 keV).

\noindent
{\bf 1249-637.} (1H 1249-637, 2E 1239.8-6246, BZ Cru)
Detected by HEAO1 (\citealt{wood1984}).
Probably a white dwarf accretor.

\noindent
{\bf 1253-761.} (1H 1253-761) Detected by HEAO1 (\citealt{wood1984}).
Probably a white dwarf accretor.

\noindent
{\bf 1255-567.} (1H 1255-567, $\mu^2$ Cru)
Detected by HEAO1 (\citealt{wood1984}).

\noindent
{\bf 1258-613.} (GX 304-1, 4U 1258-61, V850 Cen, H 1258-613, 2S 1258-613,
3A 1258-613 )
Discovered by UHURU. In \cite{z2002} classified as B0.7Ve.

\noindent
{\bf 1417-624.}  (2S 1417-624, 2S 1417-62, 4U 1416-62, 2E 1417.4-6228, 3A
1417-624, H 1417-624)
The X-ray source 2S 1417-62 was detected by SAS-3 in 1978
(\citealt{ank1980}).
Analysis of the SAS 3 observations showed an evidence of $\sim$
57 mHz pulsations (\citealt{kad81}).
Einstein and optical observations identified a Be star companion at
a distance of 1.4-11.1 kpc (\citealt{gpc84}).
From the timing analysis of BATSE
observations orbital parameters were determined
and a correlation was found between spin-up rate and
pulsed flux (\citealt{finger1996}).
Orbital period and eccentricity of the source were found to be
42.12 days and 0.446 respectively.

\noindent
{\bf J1452.8-5949.} (1SAX J1452.8-5949)
1SAX J1452.8-5949 was discovered during a BeppoSAX
galactic plane survey in 1999 (\citealt{oop1999}).
Coherent pulsations were detected with a barycentric period of
a $437.4 \pm 1.4$ s. The X-ray properties and lack of an obvious optical
counterpart are consistent with a Be star companion at a distance of
between approximately 6 and 12 kpc.
Pulse fraction is high. It was determined in the BeppoSAX band 1.8-10
keV.
Be/X-ray systems display a correlation between their spin and orbital
periods (\citealt{c86}) which in this case implies an
orbital period of $>$200 days for 1SAX~J1452.8-5949.

\noindent
{\bf J1543-568.} (XTE J1543-568)
The transient X-ray source XTE J1543-568 was discovered
by RXTE in 2000 (\citealt{icm2001}).
A subsequent pointed PCA observation revealed a pulsar with a period
of $27.12 \pm 0.02$~s. Later
the pulsar was found in earlier data from
BATSE on board the Compton Gamma-Ray Observatory.
The orbital period is $75.56 \pm 0.25$ d. The mass function and position
in the pulse period versus orbital period diagram are consistent with
XTE J1543-568 being a Be/X-ray binary. The eccentricity is less than
0.03,
so it is among the lowest for twelve Be/X-ray binaries
whose orbits have now been well measured.
This confirms the suspicion that small kick velocities of neutron
stars in HMXBs are more common for these systems than it is generally expected 
for neutron stars \citealp{icm2001,petal2004}.
There is only a lower limit for its distance. Optical component is unknown,
so \citet{icm2001} were able only to put limits V=21 for 10 kpc and V=23 for
26 kpc. The spectral class determination given by \citet{z2002} is,
probably, a misprint (see also \citet{on2001}). 
Pulse fraction (RXTE) slightly depends on energy (from 2 to 20 keV).

\noindent
{\bf 1553-542.} (2S 1553-542, 2S 1553-54, H 1553-542)
The X-ray source 2S 1553-542 was discovered during observations with
SAS 3 in 1975 (\citealt{kra1983}).
Pulse fraction was determined by SAS-3 (2-11 keV).

\noindent
{\bf 1555-552.} (1H 1555-552, LS 3417, RX J155422.2-551945, 2E
1550.3-5510, 1E 1550.4-5510)
Detected by HEAO1 (\citealt{wood1984}).

\noindent
{\bf J170006-4157.}  (AX J170006-4157, AX J1700-419, AX J1700.1-4157)
This source was discovered and observed three times between 1994 and
1997 by ASCA (\citealt{torii1999}). Significant
pulsations with P = $714.5 \pm 0.3$ s
were discovered from the third observation.
The X-ray spectrum is
described by a flat power-law function with a photon index of —0.7.
Although
the spectrum could also be fitted by thermal models, the obtained
temperature
was unphysically high. The hard spectrum suggests that the source is a
neutron
star binary pulsar similar to X Persei (4U~0352+309), but the possibility
that it is a white dwarf binary cannot be
completely excluded.
Not marked as a Be-candidate in \cite{liu2000}.
Pulse fraction in the range 0.7-10 keV was
determined from the graph in \cite{torii1999}.

\noindent
{\bf J1739-302.} (XTE J1739-302, AX J1739.1-3020)
This source was discovered during observations of the black hole
candidate 1E 1740.7-2942 with the proportional counter array (PCA)
of the Rossi X-Ray Timing Explorer (\citealt{smith1998}).
Luminosity estimated for a 2-100 keV range with an assumption, that the
source is at the Galactic center.
\cite{smith1998} tentatively identified XTE J1739-302 as a Be/NS binary
because its spectral shape is similar to that of these systems:
a gradual steepening over the 2-25 keV range.

\noindent
{\bf J1739.4-2942.} (RX J1739.4-2942)
Discovered by ROSAT (\citealt{motch1998}).
Probably identical with GRS 1736-297.

\noindent
{\bf J1744.7-2713.} (RX J1744.7-2713, HD 161103, V3892 Sgr, LS 4356)
Discovered by ROSAT (\citealt{mhd1997}).
The luminosity was estimated for the energy range 0.1-2.4 keV.
The pulse fraction was taken from paper by \cite{hwf04}. It has been
obtained by BATSE in the range 20-40 keV.

\noindent
{\bf J1749.2-2725.}  (AX J1749.2-2725)
Discovered by ASCA (\citealt{torii1998}).
Not marked as a Be-candidate in \cite{liu2000}.

\noindent
{\bf J1750-27.} (GRO J1750-27, AX J1749.1-2639)
GRO J1750-27 is the third of the transient accretion-powered
pulsars discovered using BATSE.
A single outburst from GRO J1750-27 was observed with BATSE
(see \citealt{scott1997}).
Pulsations with a 4.45~s period were discovered on 1995 July 29
from the Galactic center region as part of the BATSE all-sky pulsar
monitoring
program (\citealt{bcc97}).
An orbit with a period of 29.8 days was found by \cite{scott1997}.
Large spin-up rate,
spin period and orbital period together
suggest that accretion is occurring from a disk
and that the outburst is a ``giant'' one typical
for a Be/X-ray transient system.

\noindent
{\bf J1820.5-1434.} (AX J1820.5-1434)
This X-ray source was discovered in 1997 by  ASCA  (\citealt{kth1998}).
Pulsations with a period $\sim$ 152 s were detected in the 2-10 keV flux of
the source with
a pulsed fraction of $\sim$ 50\%. The pulse fraction is not energy
dependent.
Both timing and spectral properties of AX J1820.5-1434 are typical for
an accretion-driven X-ray pulsar. \cite{icp2000} proposed O9.5-B0Ve
star as an optical counterpart of the pulsar.

\noindent
{\bf 1843+00.} (GS 1843+00)
The transient X-ray source GS 1843+00 was discovered during
the Galactic plane scan near the Scutum region by X-ray detectors
on board the Ginga satellite (\citealt{ttp89}). Coherent pulsations
with a period of about 29.5 s were observed with a very small
peak-to-peak
amplitude of only 4 per cent of the average flux.
Pulse fraction was obtained by BATSE (20-50 keV).
Luminosity estimates are the following:
1) $2\cdot 10^{36}$~erg~s$^{-1}$ (20-200~keV, 10
kpc) (\citealt{m1999}); 2) $3\cdot 10^{37}$~erg~s$^{-1}$
(0.3-100 keV, 10 kpc) (\citealt{pss2000}).

\noindent
{\bf 1845-024.} (2S 1845-024, GS 1843-02, 4U 1850-03, 1A 1845-02, 1H
1845-024, 3A 1845-024, GRO J1849-03)
The pulsar GS 1843-02 was discovered by
Ginga in 1988 (\citealt{makino88}) during a galactic
plane scan conducted as part of a search for transient pulsars
(see \citealt{finger1999}). The same source is known as GRO J1849-03.
X-ray outbursts occur regularly every 242 days. \cite{finger1999}
presented
a pulse timing analysis that shows that the 2S 1845-024 outbursts
occur near the periastron passage. The orbit is highly eccentric
(e = $0.88 \pm 0.01$) with a period of $242.18 \pm 0.01$ days.
The orbit and transient outburst pattern strongly suggest that the pulsar
is in a binary system with a Be star. From the measured spin-up rates and
inferred luminosities \cite{finger1999} concluded that an accretion disc
is
present during outbursts.

\noindent
{\bf J1858+034.} (XTE J1858+034)
The hard X-ray transient XTE J1858+034 was discovered with the
RXTE All Sky Monitor  in 1998 (\citealt{rl98}). The spectrum was
found to be hard similar to spectra of X-ray pulsars. Observations were
made
immediately after this with the Proportional Counter Array (PCA) of the
RXTE and
regular pulsations with a period of $221.0 \pm 0.5$ s were discovered
(\citealt{tcm1998}).
The pulse profile is found to be nearly sinusoidal with a pulse fraction
of
$\sim$ 25\%. From the transient nature of this source and pulsations they
suggested
that this is a Be/X-ray binary. The position of the X-ray source
was refined by scanning the sky around the source with the PCA
(\citealt{mlc98}).
From the RXTE target of opportunity (TOO) public archival data of the
observations
of XTE J1858+034, made in 1998, \cite{pr1998} have discovered the
presence
of low frequency QPOs. Pulse fraction was obtained by RXTE (2-10 keV).

\noindent
{\bf 1936+541.} (1H 1936+541)
Detected by HEAO1 (\citealt{wood1984}).

\noindent
{\bf J1946+274.} (XTE J1946+274, GRO J1944+26, 3A 1942+274, SAX
J1945.6+2721)
Pulse fraction obtained by Indian X-ray Astronomy Experiment - IXAE (2-18
keV).
\cite{chr2002} present a data on cyclotron feature in the spectrum
of XTE~J1946+274 which corresponds to the field $\sim 3.9\cdot 10^{12}$~G.
\cite{wfc2003} propose a distance $9.5\pm 2.9$~kpc basing on a
correlation between measured spin-up rate and flux.

\noindent
{\bf J1948+32.} (GRO J1948+32, GRO J2014+34, KS 1947+300)
This transient X-ray source was discovered in 1989 during the observations
of the Cyg X-1 region by the TTM telescope aboard the Kvant module of the
Mir space station (\citealt{bgs90}). The flux recorded from it was
$70 \pm10$ mCrab in the energy range 2-27 keV.
In 1994 the BATSE monitor discovered the X-ray pulsar
GRO J1948+32 with a period of 18.7 s in the same region
(\citealt{chak1995}).
\cite{gml2004} presented results which can indicate a glitch in that
system.
Based on the behavior of the pulsation period during
the outburst of 2000-2001, they determined the parameters of the binary:
the orbital period $P_\mathrm{orb} = 40.415 \pm 0.010$~d
and the eccentricity $e = 0.033 \pm 0.013$.
The optical counterpart is a B0 Ve star.
\cite{tl2005} estimated the magnetic field strength of the pulsar $\sim
2.5\cdot 10^{13}$~G,
and the distance to the binary $d = 9.5 \pm 1.1$~kpc.
The pulse fraction depends on the source's intensity, the orbital phase
and the energy range (\citealt{tl2005}).


\noindent
{\bf 2030+375.} (EXO 2030+375, V2246 Cyg)
EXO 2030+375 was discovered in 1985 May with EXOSAT satellite during
a large outburst phase (\citealt{pws89}). This outburst was first
detected at a 1-20 keV energy band and its luminosity is close to the
Eddington limit (assuming 5 kpc distance to the source) for
a neutron star (\citealt{psf85}). The X-ray emission of the transient
pulsar EXO 2030+375 is modulated by $\sim$ 42 s pulsations and
periodic $\sim$ 46 days Type I outbursts, that are produced at each
periastron passage of the neutron star, i.e. when the pulsar interacts
with
the disk of the Be star. \citet{wfc2005} presented results of observations
of
transition to global spin-up in this source somewhen between between June
2002
and September 2003.
The source is not marked as a Be-candidate in \cite{liu2000}.
 Pulse fraction was
obtained by BATSE in the range 30-70 keV (see \citealt{hwf04}).
See a detailed description in \cite{wfc2002}.

\noindent
{\bf J2030.5+4751.} (RX J2030.5+4751, SAO 49725) Discovered by ROSAT
(see \citealt{mhd1997}).
This object is marked as a likely Be/X-ray candidate in \cite{liu2000}, but
not in many other papers.
The pointing data show that the X-ray source is relatively hard.
The $L_\mathrm{x} /L_\mathrm{bol}$ ratio is close to $3\cdot 10^{-6}$. This
is rather
strong evidence
in favor of an accreting compact object around SAO 49725
(\citealt{mhd1997}).

\noindent
{\bf J2058+42.} (GRO J2058+42, CXOU J205847.5+414637?)
GRO J2058+42 is a transient 198~s X-ray pulsar. It was discovered by BATSE
during a ``giant'' outburst in 1995 (see \citealt{wfh1998}). The pulse
period decreased from 198 to 196~s
during the 46 day outburst. BATSE observed five weak outbursts from GRO
J2058+42 that were spaced by
about 110 days. The RXTE All-Sky Monitor detected eight weak outbursts with
approximately
equal durations and intensities. GRO J2058+42 is most likely a Be/X-ray
binary that appears to produce
outbursts at periastron and apastron. No optical counterpart has been
identified to date (see however \citealt{ct96}), and no X-ray source was
present in
the error circle in archival ROSAT observations (\citealt{wfh1998}).
\cite{wwf2005} have suggested that GRO J2058+42 and CXOU J205847.5+414637
are the same source. Pulse fraction was obtained by BATSE in the range
20-70 keV
(see \citealt{hwf04} for details).

\noindent
{\bf J2103.5+4545.} (SAX J2103.5+4545)
SAX J2103.5+4545 is a transient HMXB pulsar with a $\sim 358$ s pulse
period discovered with the WFC on-board BeppoSAX during an outburst in 1997
(\citealt{hzh98}). Its orbital period of 12.68 days has been found with the RXTE during the
1999 outburst (\citealt{b2000}). The likely optical counterpart, a Be star with
a magnitude V=14.2, has been recently discovered (\citealt{rm2003}).
During the outburst in 1999 \cite{b2002} for the first time observed
with RXTE a transition from the spin-up phase to the spin-down regime,
while the X-ray flux was declining. \cite{i2004} observed a soft
spectral component (blackbody with a temperature of
1.9 keV) and a transient 22.7 s QPO during a XMM-Newton observation
performed in 2003. The pulsed fraction increases with energy
from $\sim 45\%$ at 5-40 keV to $\sim 80\%$ at 40-80 keV (\citealt{fdb2005}.

\noindent
{\bf 2138+568.} (GS 2138-56, Cep X-4, V490 Cep, 1H 2138+579, 4U 2135+57,
3A
2129+571)
The X-ray source Cep X-4 was discovered with a transient high level X-ray
flux in 1972 by OSO-7 (\citealt{ubw73}). The source was not detected
again
till 1998 when a new outburst was detected by GINGA. During these
observations
coherent 66~s pulsations were discovered revealing
an X-ray pulsar with a complex X-ray spectrum including a possible 30~keV
cyclotron absorption feature (\citealt{kkt91}, \citealt{mmk1991}).
Cep X-4 has been associated with a Be star that lies within the ROSAT
error
box. A cyclotron line was detected by \cite{mmk1991},
it corresponds to the magnetic field $B=2.3\cdot 10^{12} (1+z)$~G.
Pulse fraction strongly depends on energy and is highly variable with
time from nearly 0 up to $>80$\% (see \citealt{wfs1999b}).
The RXTE pulse fraction is decreasing with intensity.

\noindent
{\bf 2206+543.}  (3U 2208+54, 4U~2206+54, 1H 2205+538, 1A 2204+54, 3A
2206+543)
The hard X-ray source 4U 2206+54 was first detected by the Uhuru
satellite
(\citealt{gmg72}). The source
is included in the Ariel V catalogue as 3A 2206+543 (\citealt{wmf81}).
4U 2206+54 has been detected by all satellites that have pointed at it
and
has never been observed to undergo an outburst.
\cite{sfg84} used the refined position from the HEAO-1 Scanning
Modulation
Collimator to identify the optical counterpart with the early-type star
BD
+53$\degr$ 2790.
From their photometry, they estimated that the counterpart was a B0~-~2e
main
sequence star,
and therefore concluded that the system was a Be/X-ray binary.
\cite{crp00} have announced the detection of a $9.570 \pm 0.004$ d
periodicity in the X-ray lightcurve.
If this is the binary period, then it would be the shortest known for
a Be/X-ray binary --- unless the $\sim$ 1.4 d periodicity
in the optical lightcurve of RX J0050.7-7316 (\citealt{co00}).
Optical and ultraviolet spectroscopy of the optical component
BD +53$\degr$ 2790
show it to be a very peculiar
object, displaying emission in H I, He I and He II lines and variability
in
the intensity of many lines of metals (\citealt{nr01}).
Strong wind troughs in the UV resonance lines suggest a large mass loss
rate. These properties might indicate
that the star displays at the same time the Of and Oe phenomena or even
a hint of the possibility that it could be a
spectroscopic binary consisting of two massive stars in addition to the
compact object (\citealt{nr01}). With
all certainty there is
an O9.5V star in the system which is probably a mild Of
star, and which likely feeds the compact
object with its stellar wind (\citealt{nr01}).
See also recent data and discussion in \cite{cp2001}.
These authors confirm the orbital period of $\sim 9.6$ days.
This value is surprisingly short if one takes into account
long spin period of the neutron star (see fig.~2, where this system is definitely
displaced from the normal trend). Spin period was not detected
in many observations. \citet{cp2001}
discuss several possibilities
other than Be/X-ray interpretation including an accreting white dwarf.
Nearly perfect alignment between magnetic and spin axis is also a
possibility.
The high-energy spectra show clear indications of
the presence of an absorption feature at 32 keV. This
feature gives strong support to the existence of a cyclotron
resonance scattering feature, which implies a magnetic field of
$3.6\cdot 10^{12}$~G (\citealt{brn2005}). A tentative detection of a cyclotron
resonant feature in absorption was also 
earlier presented by \citet{metal2004}.

\noindent
{\bf 2214+589.} (1H 2214+589)
Detected by HEAO1 \citep{wood1984}.
This object is mentioned in \cite{liu2000} as a
Be-candidate. However, it is not mentioned in many lists of Be/X-ray
systems (for example in \citealt{z2002}). Not much is known about this
source.

\noindent
{\bf J2239.3+6116.} (3A 2237+608, SAX J2239.3+6116, SAX J2239.2+6116, 3U
2233+59, 4U 2238+60)
Discovered by BeppoSAX (see \citealt{isc2001}).
SAX J2239.3+6116 is an X-ray transient which often recurs with
a periodicity of 262 d (\citealt{ihe2000}).
Because of the Be-star nature of the likely optical counterpart the
periodicity may be
identified with the orbital period of the binary.
Pulse fraction was determined from the graph in
\cite{isc2001}
as $(I_\mathrm{max}-I_\mathrm{min})/(I_\mathrm{max}+I_\mathrm{min})$.
It corresponds to the energy range $\sim 1$~--~10~keV.
$L_\mathrm{max}$ corresponds to the distance 4.4~kpc and
the highest flux $10^{-9}$~erg~cm$^{-2}$s$^{-1}$
in the energy range 2-28~keV (\citealt{isc2001}).


\subsection{Sources in the LMC: comments to the table~\ref{tableLMC}}

\begin{table*}
\caption{Be/X-Ray Binaries and candidates in the LMC\label{tableLMC}
}
\begin{tabular}{|l|c|c|c|c|c|c|c|c|c|}
\hline

Name        & Spec. & $m_V$ & $P_\mathrm{spin}$, s & $P_\mathrm{orb}$, d &  e
 &  $L_\mathrm{max}$, & Pulse   \\
 & type    &      &                            &
 &                     &  erg s$^{-1}$ & frac., \% \\

\hline

 & & & & & & & &  \\

J0501.6-7034    & B0 Ve [49] & 14.36 [49] & & & &  $7\cdot 10^{34}$ [49] & \\

J0502.9-6626    & B0 Ve [49] & 14.42 [49] & 4.1 [49] & & &  $4\cdot 10^{37}$ [49] & \\

J0516.0-6916    & B1 V [49] & 15.0 [10] & & & & $5\cdot 10^{35}$ [10] & \\
                              & Be [50] & & & & & & & \\

J0520.5-6932    & O9 Ve [49] & 14.4 [10] & & 24.4 [51] & &  $8\cdot 10^{38}$ [51] &  \\

J0529.8-6556    & B0.5 Ve [49] & 14.81 & 69.5 [49] & & &  $2\cdot 10^{36}$ [49] & \\

053109-6609.2  & B0.7 Ve [49] & & 13.7 [49] & 24.5 [49] & &  $1\cdot 10^{37}$ [49] & 54-78 [52] \\

J0531.5-6518    & B2 Ve [49] & 16.02 [49] & & & &  $3\cdot 10^{35}$ [49] & \\

J0535.0-6700    & B0 Ve [49] & 14.87 [49] & & & &  $3\cdot 10^{35}$ [49] & \\

0535-668            & B0.5 IIIe [49] & 12.3-14.9 [10] & 0.068 [10] & 16.7 [10] & $> 0.5$ [1] &  $1\cdot 10^{39}$ [49] & \\

0544-665            & B0 Ve [49] & 15.55 [49] & & & & $1\cdot 10^{37}$ [49] & \\

0544.1-710         & B0 Ve [49] & 15.25 [49] & 96.08 [49] & & &  $2\cdot 10^{36}$ [49] & \\

\hline
\label{tableLMC}
\end{tabular}
\end{table*}

References to the table 2:\\
(49) \cite{nc02};
(50) \cite{scc99};
(51) \cite{ecg2004a};
(52) \cite{bsr98};

\noindent
{\bf J0501.6-7034.} (RX J0501.6-7034,  2E 0501.8-7038, 1E 0501.8-7036, HV
2289, CAL 9)
This Einstein and ROSAT variable source was identified with a Be star by
\cite{scf94}. Later \cite{sch96} identified this star with HV~2289, a known
variable
with a large amplitude of variability.

\noindent
{\bf J0502.9-6626.} (RX J0502.9-6626, CAL E)
The X-ray source  RX J0502.9-6626 was originally detected by the Einstein
observatory
(\citealt{cch84}) at a flux of $\sim 3\cdot 10^{36}\, \mathrm{erg~
s^{-1}}$. The source
was detected three times with the ROSAT PSPC at luminosities
$\sim 10^{35} - 10^{36}\, \mathrm{erg~s^{-1}}$ and once with the HRI
during
a bright
outburst at $4\cdot 10^{37}\, \mathrm{erg~s^{-1}}$ (\cite{scm95}). During
the outburst, pulsations at
4.0635 s were detected. The identification of this source with the Be
star
[W63b]~564 = EQ~050246.6-663032.4
(\citealt{w63}) was confirmed by \cite{scf94}.

\noindent
{\bf J0516.0-6916.} (RX J0516.0-6916)
The identification of this source with a Be-star is unclear.
In several observations the source  did not display any characteristics
of Be  behaviour, however, Schmidtke et al. (1999) classify it as a
Be-star.

\noindent
{\bf J0520.5-693.} (RX J0520.5-6932)
This X-ray source has been observed at a low X-ray luminosity
($5\cdot 10^{34}$ erg s$^{-1}$)
in early 90-s by ROSAT (\citealt{scf94}).
The light curve of the optical counterpart exhibits significant
modulation with a period of 24.5 d,
which is interpreted as the orbital period (\citealt{cnb01}).
A spectral type  O9V was proposed for the optical counterpart.
In a recent paper \cite{ecg2004a} present new optical and IR data
and archive BATSE data on the outburst.

\noindent
{\bf J0529.8-6556.} (RX J0529.8-6556, RX J0529.7-6556)
The transient X-ray source RX J0529.8-6556 was detected during one single
outburst as a 69.5-s X-ray pulsar by \cite{hdp97}, who identified it with
a relatively bright blue star showing weak $H_\alpha$ emission.

\noindent
{\bf 053109-6609.2.} (EXO 053109-6609.2, RX J0531.2-6609,  RX
J0531.2-6607,
EXO 0531.1-6609)
This source was discovered by EXOSAT during deep observations of the LMC
X-4
region in 1983 (\citealt{pbp85}). It was detected again in 1985 by the
SL2
XRT experiment.
The lack of detection in EXOSAT observations made between these dates
demonstrates the
transient nature of the source. The companion is optically identified
with
a Be star (\citealt{hdp95}).
\cite{bsr98} reported a timing analysis of the Be transient X-ray binary
EXO 053109-6609.2 in outburst
observed with BeppoSAX. The pulsed fraction is nearly constant in the
whole
energy range.
The source shows pulsations from 0.1 up to 60 keV.
In the MECS (Medium Energy Concentrator Spectrometer) pulse profile in
the
1.8-10.5 keV band the pulsed fraction
is $0.54 \pm 0.05$. In the LECS (Low Energy Concentrator Spectrometer)
pulse profile (the 0.1-1.8 keV band), the
main pulse is still evident, while the interpulse is more broadened, and
pulsed fraction is
$0.78 \pm 0.28$. The PDS (Phoswich Detection System) pulse profile (15-60
keV energy band)
still shows a double-peaked structure (pulsed fraction is $0.64 \pm
0.16$)
in phase with the previous ones.
Although the statistics is poor, the pulsed fraction does not seem to
decrease with energy (\citealt{bsr98}).

\noindent
{\bf J0531.5-6518.} (RX J0531.5-6518)
This source was detected with the ROSAT PSPC in June 1990
(\citealt{hp99}). The source is probably variable, since other pointings
failed to detect
it. The optical counterpart is probably a Be star coming back from an
extended disk-less phase
(\citealt{nc02}).

\noindent
{\bf J0535.0-6700.} (RX J0535.0-6700)
This source was observed by the ROSAT PSPC at a luminosity
$\sim 3\cdot 10^{35} \mathrm{erg~s^{-1}}$ (\citealt{hp99}). Its
positional coincidence with an optically variable star in the LMC (RGC28
in \citealt{rgc88}) is very good.
RGC28 is an early-type Be star and likely it is the optical counterpart
to
RX~J0535.0-6700 (\citealt{nc02}). The star displays periodic variability
in its I-band lightcurve at 241 d, which \cite{rgc88} originally believed
to be the period of a Mira variable. \cite{hp99} suggested that this
variability
can be related to the orbital period.

\noindent
{\bf 0535-668.} (RX J0535.6-6651, 1A 0538-66, 1A 0535-66)
This source was discovered by the Ariel 5 satellite in June 1977, during
outburst in which the flux peaked at $\sim 9\cdot 10^{38} \mathrm{erg~
s^{-1}}$ (\citealt{wc78}). When active, 1A 0535-66 displays very bright
short
X-ray outbursts separated by 16.6 days, which is believed to be the
orbital period. The optical counterpart experiences drastic changes in the
spectrum, with the
appearance of strong P-Cygni-like emission lines, and brightening by more
than 2 mag in the
$V$ band (\citealt{cbd83}).
The Be star has a $V$ magnitude of $\sim 14.8$ during the X-ray quiescent
periods. The magnitude reaches a peak of 12.5 mag during the X-ray
outbursts.
Detection of a 69-ms pulsation in the X-ray signal has been reported only
once (\citealt{sbe82}).
Further X-ray observations of outbursts were made by \cite{sss80} using
the HEAO 1 satellite. The X-ray outbursts were found to last up to at least
14
days or to be as short as a few hours. 1A~0535-66 in its largest
outbursts (\citealt{sss80}) has luminosity around $10^{39}\, \mathrm{erg
s^{-1}}$. ROSAT (\citealt{mh93}) and ASCA observations (\citealt{csc95})
have
revealed low-level outbursts with luminosities of $4\cdot 10^{37}\,
\mathrm{erg ~s^{-1}}$
and $2\cdot 10^{37}\, \mathrm{erg ~s^{-1}}$ in the two ROSAT observations
and $\sim 5.5\cdot 10^{36}\, \mathrm{erg~s^{-1}}$ in the ASCA
observation.
Due to the low count rate and sampling frequency it was not possible to
determine whether the 69 ms pulsations were present in the data.
The ratio of $L_\mathrm{max}$ to $L_\mathrm{min}$ in soft X-rays is 
$>1000$. \cite{a2001} reported the discovery of 421~day periodicity.

\noindent
{\bf 0544-665.} (H 0544-665, H 0544-66)
This source was discovered with the HEAO-1 scanning modulation collimator
by \cite{jbd79}.
The brightest object within the X-ray error circle was found to be a
variable B0-1 V star (\citealt{kte83}) but no emission lines have been
observed in its
spectrum to identify it as a Be star. \cite{kte83} published photometry
which showed a
negative correlation between optical magnitudes and color indices, typical
of Be stars whose
variability is due to variations in the circumstellar disc. \cite{scb99}
suggested that
the object may be a Be star in the state of low activity.

\noindent
{\bf J0544.1-710.} (RX J0544.1-7100, AX J0544.1-7100, AX J0548-704,
1WGA J0544.1-7100, 1SAX J0544.1-7100)
This source is a transient X-ray pulsar ($P=96$ s) with hard
X-ray spectrum observed by BeppoSAX (\citealt{cetal1998})
and by ROSAT in the LMC (\citealt{hp99}). The
observations of the optical
counterpart were presented by \cite{cnb01}, who found it to display large
variability in
the $I$-band lightcurve and $H_\alpha$ in emission. An approximate spectral
type of B0~Ve was
proposed.


\subsection{Sources in the SMC: comments to the table~\ref{tableSMC1} and
\ref{tableSMC2}}

\begin{table*}
\caption{Be/X-Ray Binaries and candidates in the SMC\label{tableSMC}
(beginning)}
\begin{tabular}{|l|c|c|c|c|c|c|c|c|c|}
\hline

Name        & Spec. & $m_V$ & $P_\mathrm{spin}$, s & $P_\mathrm{orb}$, d
 &  $L_\mathrm{max}$, & Pulse   \\
 & type          &                            &
 &                     &  erg s$^{-1}$ & frac., \% \\

\hline

 & & & & & & & &  \\

J0032.9-7348 & Be & & & &  $1.3\cdot 10^{37}$ [5] & \\

J0045.6-7313 & Be? & & & &  $1.2\cdot 10^{35}$ [5] & \\

J0047.3-7312 & Be? & 15.85 [53] & 263 [54] & 48.8 [94] &  $1.8\cdot 10^{36}$ [54] & 47 [55]\\

J0048.2-7309 & Be? & & & &  $5.2\cdot 10^{35}$ [56] & \\

J0048.5-7302 & Be? & & & &  $3.0\cdot 10^{35}$ [54] & \\

J0049-729 & Be & 16.92 [5] & 74.67 [57] & 33.3 [96] &  $7.5\cdot 10^{36}$ [58] & 70 [58] \\

J0049.2-7311 & Be? & & & &   $2.0\cdot 10^{35}$ [54] & \\

J0049-732 & Be? & 16.51 [53] & 9.1320 [59] & 91.5 [60] &   $4.1\cdot 10^{35}$ [59] & \\

J0049.5-7331 & Be? & & & &  $5.1\cdot 10^{35}$ [56] & \\

J0049.7-7323  & Be & 15.22 [53] & 755.5 [61] & 394 [53] &  $7.7\cdot 10^{35}$ [54] & \\

J0050.7-7316  & B0-B0.5 Ve [62] & 15.4 [62] & 323.1 [63] & &  $1.8\cdot 10^{36}$ [5] & 41 [64] \\
(DZ Tuc)        & & & & & & & & \\

J0050.7-7332 & Be? & & & &  $2.4\cdot 10^{34}$ [5] & \\

J0050.9-7310 & Be? & & & &  $4.5\cdot 10^{35}$ [54] & \\

J0051-722  & Be & 15.06 [53] & 91.11 [5] & 88.4 [53] &  $2.9\cdot 10^{37}$ [5] & \\

J0051.3-7250 & Be? & & & &  $3.6\cdot 10^{34}$ [54] & \\

J0051-727 & & & 293 [65] & &  $1.7\cdot 10^{37}$ [65] &  \\

J0050-732 \#1 & & & 16.6 [66] & 189 [68] &  $3.7\cdot 10^{36}$ [66] & \\

J0050-732 \#2 & & & 25.5 [66] & &  $3.0\cdot 10^{36}$ [66] &  \\

J0051.8-7310  & Be & 14.45 [53] & 172.4 [67] & 147 [68] &  $5.6\cdot 10^{36}$ [68] & \\

J0051.8-7231  & Be & 14.87 [53] & 8.9 [54] & 185 [53] &  $1.4\cdot 10^{36}$ [5] & 25 [69] \\

WW 26 & Be? & & & &  $6.0\cdot 10^{34}$ [5] & \\

0050-727   & O9 V-IIIe [10] & 14.91 [53] & 7.78 [64] & 44.6 [53] &  $6\cdot 10^{37}$ [70] & 27 [64] \\
(SMC X-3) & & & & & & & & \\

J0052.1-7319  & O9.5 IIIe [71] & 14.67 [53] & 15.3 [5] & &  $1.3\cdot 10^{37}$ [5] & \\

J0052-725 & Be? & 15.02 [53] & 82.46 [64] & &  $3.4\cdot 10^{36}$ [60] & 19-42 [64] \\

J0052-723 & B0V-B1Ve [72] & 15.8 [72] & 4.78 [54] & &  $7.2\cdot 10^{37}$ [72] &  \\

0051.1-7304  & Be & 14.28 [5] & & &  $1.6\cdot 10^{35}$ [5] & \\

J0052.9-7158  & Be & 15.53 [53] & 167.8 [73] & &  $2.0\cdot 10^{37}$ [5] & 44 [73] \\

J005323.8-722715 & Be? & 16.19 [53] & 138.04 [64] & 125 [64] &  $1.18\cdot 10^{35}$ [64] & 59 [64] \\

XTE SMC 95 & & & 95 [74] & &  $2\cdot 10^{37}$ [74] & \\

J0055-727 & Be? & & 18.4 [75] & 34.8 [76] &  $2.6\cdot 10^{37}$ [75] & \\

J0053.8-7226  & Be & 14.72 [53] & 46.63 [54] & 139 [77] &  $7.4\cdot 10^{36}$ [5] & 25 [77] \\

0053-739        & B1.5 Ve [78] & 16.64 [53] & 2.37 [54] & &  $4.7\cdot 10^{38}$ [68] & \\
(SMC X-2)      & & & & & & & & \\

J0054.5-7228 & Be? & & & &  $1.5\cdot 10^{36}$ [5] & \\

J0054.8-7244 & Be? & 14.99 [53] & 503.5 [64] & 268 [95] &  $5.5\cdot 10^{35}$ [64] & 63 [64] \\

J0054.9-7226   & B0-B1 III-Ve [10] & 15.28 [5] & 59.07 [5] & 65 [79] &  $4.3\cdot 10^{37}$ [68] \\

J005517.9-723853 & Be? & 16.01 [53] & 701.7 [80] & &  $4\cdot 10^{35}$ [80] & \\

J0055.4-7210 & Be? & 16.78 [53] & 34.08 [64] & &  $1.1\cdot 10^{35}$ [64] & 57 [64] \\

0054.4-7237 & Be? & 16.51 [53] & 140.1 [54]  & &  $4.0\cdot 10^{34}$ [5] & \\

J0057.4-7325 & Be? & & 101.45 [81] & &  $1.2\cdot 10^{36}$ [82] & \\

J005736.2-721934 & Be? & 15.97 [53] & 562 [77] & 95.3 [53] &  $1.2\cdot 10^{36}$ [82] & 73 [64] \\

J0057.8-7202  & Be? & 15.65 [53] & 281.1 [83] & &   $1.6\cdot 10^{36}$ [5] & \\

J0057.8-7207 & Be? & & 152.3 [54] & &  $4.3\cdot 10^{35}$ [5] & \\

J0057.9-7156 & Be? & & & &  $5.7\cdot 10^{34}$ [5] & \\

J0058.2-7231 & B2-3 Ve [84] & 14.90 [84] & & 59.72 [85] &  $2.1\cdot 10^{35}$ [5] & \\

J0059.2-7138 & B0 III-Ve [68] & 14.08 [5] & 2.763 [5] & &  $5.0\cdot 10^{37}$ [5] & 37 [86] \\

J0059.3-7223 & Be? & 14.83 [53] & 202 [53] & &  $3.2\cdot 10^{35}$ [56] & \\

J010030.2-722035 & Be? & & & &  $2.6\cdot 10^{34}$ [83] & \\

J0101.0-7206   & Be & 15.72 [53] & 304.49 [87] & &  $1.3\cdot 10^{36}$ [5] &  \\

J0101.3-7211 & Be & 15.49 [53] & 455 [54] & &  $7.3\cdot 10^{35}$ [13] & \\

J0101.6-7204  & Be? & & & &  $3.8\cdot 10^{35}$ [5] & \\

J0101.8-7223  & Be? & & & &  $2.2\cdot 10^{35}$ [5] & \\

J0103-728 & & & 6.85 [54] & &  $3.8\cdot 10^{37}$ [88] & \\

J0103-722 & O9-B1 III-Ve [10] & 14.80 [5] & 345.2 [5] & &  $1.5\cdot 10^{36}$ [5] & \\

J0103.6-7201 & O5 Ve [97] & 14.6 [97] & 1323 [97] & &  $6.4\cdot 10^{36}$ [97] & \\

J0104.1-7244  & Be? & & & &  $3.8\cdot 10^{34}$ [5] & \\

J0104.5-7221 & Be? & & & &  $4.8\cdot 10^{34}$ [5] & \\

J0105-722 & Be? & 15.63 [53] & 3.343 [5] & &  $1.5\cdot 10^{35}$ [5] & \\

J0105.9-7203 & Be? & & & &  $6.5\cdot 10^{34}$ [5] & \\

J0106.2-7205 & B2-5 III-Ve [10] & 16.7 [10] & & &  $5\cdot 10^{34}$ [89] & \\

\hline
\label{tableSMC1}
\end{tabular}
\end{table*}

\begin{table*}
\caption{Be/X-Ray Binaries and candidates in the SMC\label{tableSMC}
(continue)}
\begin{tabular}{|l|c|c|c|c|c|c|c|c|c|}
\hline

Name        & Spec. & $m_V$ & $P_\mathrm{spin}$, s & $P_\mathrm{orb}$, d
 &  $L_\mathrm{max}$ & Pulse   \\
 & type          &                            &
 &                     &  erg s$^{-1}$ & frac., \% \\

\hline

 & & & & & & & &  \\

J0107.1-7235 & Be? & & & &  $2.3\cdot 10^{34}$ [5] & \\

0107-750 & Be & 17 [10] & & &  $4.3\cdot 10^{35}$ [56] & \\

J0111.2-7317 & B0.5-B1Ve [18] & 15.52 [53] & 31.03 [5] & &  $2.0\cdot 10^{38}$ [5] & 45 [71] \\

J0117.6-7330 & B0.5 IIIe [90] & 14.18 [53] & 22.07 [5] & &  $1.2\cdot 10^{38}$ [5] & 11.3 [91] \\

J0119.6-7330 & Be? & & & &  $1.5\cdot 10^{34}$ [5] & \\

J0119-731 & Be? & & 2.165 [54] & &  $6.3\cdot 10^{36}$ [92] & \\

SXP46.4 & & & 46.4 [54] & &  & \\

SXP89.0 & & & 89 [54] & &  & \\

XTE SMC144s & & & 144.1 [54] & 61.2 [93] &  & \\

SXP165 & & & 164.7 [54] & &  & \\

\hline
\label{tableSMC2}
\end{tabular}
\end{table*}

We often refer to the well known ``A New Catalogue of $H_\alpha$ Emission
Line Stars and Small Nebulae in the Small Magellanic Cloud''
by \cite{ma1993}; below we use an abbreviation MA93 for this catalogue in the
comments for particular objects.

References for the tables 3 and 4:\\
(53) \cite{ceg2004};
(54) \cite{hp2004};
(55) \cite{mlm2004};
(56) \cite{yit03};
(57) \cite{yk98a};
(58) \cite{yit99};
(59) \cite{uyi00a};
(60) \cite{scl2004a};
(61) \cite{uyi00b};
(62) \cite{chl02};
(63) \cite{iyt99};
(64) \cite{ecg2004b};
(65) \cite{cmc2004};
(66) \cite{lmp2002};
(67) \cite{yti00};
(68) \cite{lcc04};
(69) \cite{isa95};
(70) \cite{ljc77};
(71) \cite{cnc01};
(72) \cite{lcc03};
(73) \citealt{ytk01b};
(74) \cite{lcp2002};
(75) \cite{cmc2003};
(76) \cite{cmm2004};
(77) \cite{l98};
(78) \cite{mmt79};
(79) \cite{lwc99};
(80) \cite{hps04};
(81) \cite{ytk00};
(82) \cite{tky00};
(83) \cite{sph03};
(84) \cite{ec03};
(85) \cite{scl2004b};
(86) \cite{kyk00};
(87) \cite{mfl03};
(88) \cite{cmm2003};
(89) \cite{hs94};
(90) \cite{s99};
(91) \cite{mfh98};
(92) \cite{cmm2003c};
(93) \cite{clm2003};
(94) \cite{ecg2005a};
(95) \cite{ecg2005b};
(96) \cite{sc2004};
(97) \cite{hp05}

\noindent
{\bf J0032.9-7348.} (RX J0032.9-7348)
This source was discovered by \cite{kp96}
in ROSAT pointed observations made in 1992 December and 1993 April.
\cite{scb99} identified two Be stars within PSPC error circle of
RX~J0032.9-7348.

\noindent
{\bf J0045.6-7313.} (RX J0045.6-7313)
This source was detected once in the 0.9-2.0 keV band of the ROSAT PSPC.
An emission-line object in the error circle suggests a Be/X-ray binary
(\citealt{hs00}).

\noindent
{\bf J0047.3-7312.} (RX J0047.3-7312, 2E 0045.6-7328, XMMU
J004723.7-731226, SXP 264, AX J0047.3-7312?)
\cite{hs00} proposed RX J0047.3-7312 as a Be/X-ray binary candidate
because this source exhibits a flux variation with a factor of 9 and has an
emission-line object 172 in MA93 as a counterpart.
A probable binary period of $48.8 \pm 0.6$ days has been detected in
observations of
the optical counterpart to this X-ray source (\citealt{ecg2005a}).
The relationship between this orbital period and the pulse
period of 264~s is within the normal variance found in the Corbet
diagram (\citealt{c1984}).

\noindent
{\bf J0048.2-7309.} (AX J0048.2-7309)
AX J0048.2-7309 was detected in two ASCA observations and shows a hard
spectrum and a flux variability with a factor of $\sim 5$ (\citealt{yit03}).
An emission-line object, No. 215 in MA93, was found
in the error circle of AX J0048.2-7309.
Data suggest that this source is a Be/X-ray binary.

\noindent
{\bf J0048.5-7302.} (RX J0048.5-7302, XMMU J004834.5-730230)
The emission-line object 238 in MA93 is the brightest optical object in
the error circle of RX J0048.5-7302 (\citealt{hs00}).  A Be/X-ray binary
interpretation is
suggested by \cite{hs00}.

\noindent
{\bf J0049-729.} (AX J0049-729, AX J0049-728, RX J0049.0-7250, RX
J0049.1-7250,  XTE J0049-729)
This source was discovered with ROSAT (\citealt{kp96}) in pointed
observations. \cite{yk98a} reported X-ray pulsations in ASCA data
of this source. The X-ray flux in the band
0.7-10 keV was  $1.2 \cdot 10^{-11}\ \mathrm{erg~cm^{-2} s^{-1}}$, with
sinusoidal pulse modulation. \cite{kp98} suggested the highly variable
source RX J0049.1-7250 as a counterpart. \cite{scb99} identified two
Be stars, one only $3\arcsec$ from the X-ray position and one just
outside the error circle given by  \cite{kp96}.
\cite{yit99} reported on the results of two ASCA observations of this
X-ray source. The pulse fraction was $ \sim 70\%$ independent of the X-ray energy.
Using MACHO and OGLE-II data, \cite{sc2004} found a strong periodicity at $P=33.3$
days, which is likely the orbital period, in good agreement with the relation
between orbital and pulse periods first recognized by \cite{c1984}.

\noindent
{\bf J0049.2-7311.} (RX J0049.2-7311, XMMU J004913.8-731136)
The position of $H_{\alpha}$ bright object coincides with this X-ray
source.
\cite{ceg2004} proposed this object as a more likely counterpart to the
ASCA
SXP9.13 pulsar. However, other authors have proposed  RX J0049.4-7310, as
the
correct identification for SXP9.13 pulsar (\citealt{fhp00};
\citealt{scl2004a}).

\noindent
{\bf J0049-732.} (AX J0049-732,  RX J0049.4-7310)
This source was discovered as an X-ray pulsar by \cite{iyk98} with ASCA.
The X-ray flux at 2-10 keV was about $8 \cdot 10^{-13}\
\mathrm{erg~cm^{-2}
s^{-1}}$. A more likely scenario for  AX J0049-732 is either a Be/X-ray
binary or
an anomalous X-ray pulsar. Direct information to distinguish these
two possibilities can be obtained by measuring the pulse period
derivative and its orbital modulation.
Two sources, No. 427 and No. 430, in the ROSAT PSPC catalogue of
\cite{hfp00} are possible counterparts of  AX J0049-732.
\cite{fph00} searched for optical counterparts of these ROSAT sources,
and found an emission line object, possibly a Be star, at the position of
source No. 427, but found no counterpart for source No. 430. Hence, they
suggest that source No. 427 is more likely to be a counterpart of  AX
J0049-732. However, the angular separation of these sources
of $1.\arcmin 43$ is significantly larger than the ASCA error radius.
And \cite{uyi00a} propose that No. 430 is a more likely counterpart.
\cite{scl2004a}  found an orbital period of 91.5 days for RX J0049.4-7310
in the MACHO data.

\noindent
{\bf J0049.5-7331.} (RX J0049.5-7331, AX J0049.5-7330)
This source is most likely identified with the emission-line
object 302 in MA93 (\citealt{hs00}).

\noindent
{\bf J0049.7-7323.} (AX J0049.4-7323, AX J0049.5-7323, RX J0049.7-7323)
This X-ray source has been detected 5 times to date, 3 times by the ASCA
observatory (\citealt{yiu00}) and 2 times by the RossiXTE spacecraft
(\citealt{lcc04}).
\cite{uyi00b} reported an ASCA observation which revealed coherent
pulsations of period $755.5 \pm 0.6$ s from
a new source in the Small Magellanic Cloud.
The spectrum was characterized by a flat power-law function
with photon index 0.7 and
X-ray flux $1.1\cdot 10^{-12}\ \mathrm{erg~cm^{-2} s^{-1}}$ (0.7-10 keV).
They noted that the possible Be/X-ray binary RX~J0049.7-7323
(\citealt{hs00}) was located within the ASCA error region.
\cite{ec03} reported on the spectroscopic and photometric analysis of
possible optical counterparts to AX J0049.4-7323. They detected strong
$H_\alpha$ emission from the optical source identified with RX
J0049.7-7323
within error circle
for AX J0049.4-7323 and concluded that these are one and the
same object.
They noted that the profile of the curve exhibits a distinct double peak.
This is consistent with Doppler effects which would be expected from a
circumstellar disc viewed in the plane of rotation. There is also
definite
V/R asymmetry between the peaks. It is a
compelling evidence for the presence of a Be star.
\cite{cs03} analyzed the long term light curve of the optical counterpart
obtained from the MACHO date base. They showed that the optical object
exhibited outbursts every 394~d which they proposed to be the orbital
period
of the system. They also showed the presence of a quasi-periodic
modulation
with a period $\sim$ 11d which they associated with the rotation of the
Be
star disk.
The phase of two RXTE detections is exactly syncronised with the
ephemeris
derived from the optical outbursts. Therefore, as \cite{ce04} concluded,
the period
of 394~d can represent the binary period of a
system with X-ray outbursts
syncronised
with the periastron passage of the neutron star.

\noindent
{\bf J0050.7-7316.} (DZ Tuc, AX J0051-732, RX J0050.6-7315, RX
J0050.7-7316, AX J0051-733, RX J0050.8-7316)
This X-ray source was detected in Einstein IPC, ROSAT PSPC and HRI
archival
data and 18 year
history shows flux variations by at least a factor of 10
(\citealt{iyt99}).
The source was reported as a 323 s pulsar by \cite{yk98b} and
\cite{iyt99}.
Subsequently \cite{c98} identified a 0.7 d optically variable object
within
the ASCA X-ray error circle.
Long term optical data from over 7 years revealed both a 1.4d modulation
and an unusually rapid change in this possible binary period
(\citealt{chl02}).
The system was discussed in the context of
being a normal high mass
X-ray binary by \cite{co00}
who presented some early OGLE data on the object identified by \cite{c98}
and modeled the system
parameters. Coe \& Orosz identified several problems with understanding
this system, primarily that if it was
a binary then its true period would be 1.4 d and it would be an extremely
compact system.
In addition, the combination of the pulse period and such a binary period
violates the Corbet relationship for such systems
(\citealt{c86}). \cite{rl1998} calculated the critical orbital period for
the existence of a Be+X-ray pulsar
binary, which is $\sim 10-20$~d. They proposed an explanation for the
lack of Be stars  with accreting
neutron star as companions
with orbital periods less than 10 days as caused by synchronization of
Be star during its evolution.
\cite{chl02} reported on extensive new data sets from both OGLE and
MACHO, as well as on detailed
photometric study of the field. Their results reveal many complex
observational features that are hard to explain
in the traditional Be/X-ray binary model.

\noindent
{\bf J0050.7-7332.} (RX J0050.7-7332)
This source was only once detected by the ROSAT PSPC.
The emission-line object in the error circle suggests a Be/X-ray
binary identification (\citealt{hs00}).

\noindent
{\bf J0050.9-7310.} (RX J0050.9-7310, RX J0050.8-7310, XMMU
J005057.6-731007)
This source is most likely identified with the emission-line
object 414 in MA93, suggesting a Be/X-ray binary (\citealt{hs00}).

\noindent
{\bf J0051-722.} (AX J0051-722,  RX J0051.3-7216)
This source was at first detected as a 91.12 s pulsar in RXTE
observations
(\citealt{cml98}). Although it was initially confused with the nearby 46~s
pulsar 1WGA J0053.8-7226
(\citealt{bcs98}). \cite{scb99}  estimated the magnitude of the optical
component (Be star) as $V
\sim 15$ from Digitized Sky Survey images. The spacing of flares observed
from
AX J0051-722 suggests an orbital period of about 120 days
(\citealt{isc98}).
\cite{scl2004a} found an optical period of 88.25 days using MACHO data.

\noindent
{\bf J0051.3-7250.} (RX J0051.3-7250)
Two close emission-line objects are found near this source, suggesting
RX J0051.3-7250 as Be/X-ray binary,

\noindent
{\bf J0051-727.} (XTE J0051-727)
\cite{cmc2004} have detected this new transient X-ray pulsar in the
direction of
the SMC with the RXTE Proportional Counter Array.
The object showed the $1.6-1.7$ mCrab flux in the 2-10 keV band.

\noindent
{\bf J0050-732\#1.} (XTE J0050-732\#1)
This source was discovered by \cite{lmp2002} from
archival data of RXTE.

\noindent
{\bf J0050-732\#2.} (XTE J0050-732\#2)
This source was discovered by \cite{lmp2002} from
archival data of RXTE.

\noindent
{\bf J0051.8-7310.} (2E 0050.2-7326, RX J0051.8-7310, AX J0051.6-7311, RX
J0051.9-7311, XMMU J005152.2-731033)
This X-ray source was detected by \cite{csm97} during ROSAT HRI
observations
of Einstein IPC source 25 and identified with a Be star by \cite{scc99}.

\noindent
{\bf J0051.8-7231.}  (2E 0050.1-7247, RX J0051.8-7231, 1E 0050.1-7247,
1WGA J0051.8-7231)
2E 0050.1-7247 was discovered in Einstein observations. The X-ray
luminosity, time variability and hard spectrum led \cite{kp96} to suggest
a Be/X-ray binary nature for the source.
\cite{isa95} discovered 8.9 s X-ray pulsations in 2E 0050.1-7247
during a systematic search for periodic signals in a sample of ROSAT PSPC
light curves. The signal had a nearly sinusoidal shape with a 25-percent
pulsed
fraction. The source was detected several times between 1979 and 1993 at
luminosity
levels ranging from $5 \cdot 10^{34} \mathrm{erg~s^{-1}}$ up to $1.4\cdot
10^{36}
\mathrm{erg~s^{-1}}$ with both the Einstein IPC and ROSAT PSPC.
The X-ray energy spectrum is consistent with a power-law spectrum that
steepens as the source luminosity decreases. \cite{isa97} revealed a
pronounced
$H_\alpha$ activity from at least two B stars in the X-ray error circles.
These results strongly
suggest that the X-ray pulsar 2E 0050.1-7247 is in a Be-type massive
binary.
\cite{ceg2004} have proposed an orbital period of $185 \pm 4$ days from the
red light data.

\noindent
{\bf WW 26.} (WW 26)
\cite{hs00} suggested a Be/X-ray binary nature for this object.
They have found two emission-line objects 521 and 487 from MA93 near
this source.

\noindent
{\bf 0050-727.} (SMC X-3, H 0050-727, 2S 0050-727, 3A 0049-726, 1H
0054-729, H 0048-731, 1XRS 00503-727)
SMC X-3 was detected by \cite{ljc77} with SAS 3.
This long-known X-ray  source was not detected by the ROSAT PSPC. But it
is
included in the HRI catalogue. \cite{mlt97} reported the detection with
the RXTE PCA
of an outburst from the X-ray transient SMC X-3 and the discovery of a
period of $92 \pm 1.5$ s with a complex pulse profile.
The Be star counterpart corresponds to object 531 in MA93.

\noindent
{\bf J0052.1-7319.} (1E 0050.3-7335, 2E 0050.4-7335, RX J0052.1-7319)
The X-ray transient RX J0052.1-7319 was discovered by \cite{lpm99} with
the analysis of ROSAT HRI and BATSE data. The object showed a period of
15.3 s
(\citealt{k99a}; \citealt{k99b})
and a flux in the 0.1-2 keV band of  $2.6\cdot 10^{-11}\, \mathrm{erg~
cm^{-2} s^{-1}}$. \cite{cnc01} reported on the discovery and confirmation
of the optical
counterpart of this transient X-ray pulsar. They found a $V = 14.6$
O9.5IIIe star (a classification as
a B0Ve star is also possible since the luminosity class depends on the
uncertainty
on the adopted reddening).

\noindent
{\bf J0052-725.} (XTE J0052-725)
This X-ray pulsar was originally detected by RXTE in 2002
(\citealt{cmm2002}).
Timing analysis revealed a period of $82.46 \pm 0.18$ s at a confidence
level of $> 99\%$.
The lower energy band (0.3-2.5 keV) contained about 60\% of the photons but
had a pulsed
fraction of only $28\% \pm 2\%$ as compared to $42\% \pm 3\%$ in the higher
energy band (2.5-10 keV).
This source has been identified with the optical counterpart MACS
J0052-726\#004 (\citealt{tbs96}).

\noindent
{\bf J0052-723.} (XTE J0052-723)
\cite{cmm01} discovered this transient X-ray pulsar in the direction of
the Small Magellanic Cloud from RXTE PCA
observations made on 2000 December 27 and 2001 January 5.  Pulsations
were seen with a period of $4.782 \pm 0.001$ s and with a double-peaked
pulse
profile. Spectroscopy of selected optical candidates (\citealt{lcc03})
has identified the probable counterpart which is a B0V-B1Ve SMC member
exhibiting a
strong, double peaked $H_\alpha$ emission line.

\noindent
{\bf 0051.1-7304.} (2E~0051.1-7304, AzV~138)
This source is listed as entry 31 in the Einstein IPC catalogue 
\citep{ww92}.
The Be star AzV 138 (\citealt{gh85}) was proposed as an optical
counterpart for 2E~0051.1-7304.
2E~0051.1-7304 was not detected in ROSAT observations.

\noindent
{\bf J0052.9-7158.} (2E 0051.1-7214, RX J0052.9-7158,  XTE J0054-720, AX
J0052.9-7157)
This source was detected as an X-ray transient by \cite{csm97} during
ROSAT
HRI observations of Einstein IPC source 32. The strong variability and the
hard X-ray
spectrum imply a Be/X-ray transient consistent with the suggested Be star
counterpart
(\citealt{scc99}). The X-ray source was detected by ROSAT and is located
near the edge of
the error circle of XTE J0054-720. The transient pulsar XTE J0054-720 with
spin period $\sim 169$
s was discovered with RXTE (\citealt{lmw98}). \cite{yit03} detected
coherent pulsations with
167.8 s period from AX J0052.9-7157 and determined its position accurately.
They found
that AX J0052.9-7157 is located within the error circle of  XTE J0054-720
and has a variable
Be/X-ray binary,  RX J0052.9-7158, as a counterpart. From the nearly equal
pulse period and the positional
coincidence, they concluded that the ASCA, ROSAT, and RXTE sources are
identical.
The pulsed fraction, defined as (pulsed flux)/(total flux) without
background, is
44\% in 2.0-7.0 keV (\citealt{ytk01b}).

\noindent
{\bf J005323.8-722715.} (CXOU J005323.8-722715, RX J0053.5-7227)
A precise ROSAT HRI position coincident with the
emission-line star 667 in MA93 (it is the brightest object
in the  error circle)
makes RX J0053.4-7227 a likely Be/X-ray binary (\citealt{hs00}).
The position of this pulsar is coincident also with MACHO
object 207.16202.50. The latter shows an evidence of a period
of $125 \pm 1.5$ days. This period
would be consistent with that predicted from the Corbet
diagram (\citealt{c86}) for a 138s Be/X-ray pulsar.

\noindent
{\bf XTE SMC 95.} The source has been revealed during RXTE observations
of the Small Magellanic Cloud. The pulsar was detected in three
Proportional
Counter Array (PCA) observations
during an outburst (\citealt{lcp2002}). The source is proposed to be a
Be/neutron star system on the basis
of its pulsations, transient nature and characteristically hard X-ray
spectrum. The 2-10 keV X-ray luminosity
implied by observations is $\ga 2\cdot 10^{37}\ \mathrm{erg~s^{-1}}$.

\noindent
{\bf J0055-727.} (XTE J0055-727)
This source was detected with the RXTE PCA
(\citealt{cmc2003}).
Regular monitoring of the Small Magellanic Cloud with the RXTE PCA
has revealed a periodicity of
34.8 days in the pulsed flux from this X-ray pulsar
(\citealt{cmm2004}).
The regular nature of outbursts strongly suggests that they show the
orbital period of this system. The combination
of pulse and orbital periods is consistent with XTE J0055-727 being a Be
star system.
\cite{cmc2003} noted the presence of the emission line objects AzV164 and
829 in MA93 close to the center of the error box of this source.

\noindent
{\bf J0053.8-7226.} (RX J0053.9-7226, 1WGA J0053.9-7226, 1E 0052.1-7242,
2E 0052.1-7242, RX J0053.8-7226,  1WGA J0053.8-7226,  XTE J0053-724)
This object was  serendipitously discovered as an X-ray source in the SMC
in the ROSAT PSPC archive and also was observed by the Einstein IPC.
Its X-ray properties, namely the hard X-ray spectrum,
flux variability and column density indicate a hard, transient source
with a luminosity of $3.8\cdot 10^{35}\ \mathrm{erg~s^{-1}}$
(\citealt{bcs01}). XTE and ASCA
observations have confirmed the source to be an X-ray pulsar, with a 46 s
spin period.
Optical observations (\citealt{bcs01}) revealed two possible counterparts
to
this source. Both exhibit strong $H_\alpha$ and weaker $H_\beta$ emission.
Optical colors indicate that both objects are Be stars.
The transient X-ray system XTE J0053-724 was also detected
in one observation by RXTE. Pulsations of $46.6 \pm 0.1$ s were observed
with a pulse fraction
about 25\% (\citealt{l98}). \cite{l98} suggested a possible orbital
period of this
Be/X-ray system about 139 days which is determined from the periodicity
of X-ray
outbursts.

\noindent
{\bf 0053-739.} (SMC X-2, 3A 0042-738, H 0052-739, 2S 0052-739,  H
0053-739, RX J0054.5-7340)
SMC X-2 was one of the first three X-ray sources which were discovered in
the SMC (\citealt{cdl78}). It was also detected in the HEAO 1 A-2
experiment
(\citealt{mbh79}), but not in the Einstein IPC survey (\citealt{sm81}). In
ROSAT
observations this transient source was detected only once (\citealt{kp96}).
It is
thought to be a Be/X-ray binary, since a Be star was found as its optical
counterpart (\citealt{mmt79}). In early 2000, the RXTE All-Sky Monitor
detected an
outburst at the position of SMC X-2 (\citealt{cmc01}) and a pulse period of
$2.374
\pm 0.007$ s was determined (\citealt{cm00}; \citealt{tky00}). The
source was in low luminosity state during the XMM-Newton observation
(\citealt{sph03}). In order to estimate the flux upper limit \cite{sph03}
used spectral
parameters derived by \cite{ytk01a} from the ASCA spectrum during the
outburst.
They obtained an upper limit for the un-absorbed flux of
$1.5\cdot 10^{-14}\, \mathrm{erg~cm^{-2} s^{-1}}$, corresponding
to $L_\mathrm{x} = 6.5\cdot 10^{33}\ \mathrm{erg~s^{-1}}$ ($0.3-10.0$
keV).

\noindent
{\bf  J0054.5-7228.} (RX J0054.5-7228)
\cite{hs00} have found six emission-line objects from MA93
as possible counterparts to this X-ray source. It is therefore
a likely Be/X-ray binary but the optical counterpart remains
ambiguous.

\noindent
{\bf J0054.8-7244.} (AX J0054.8-7244, RX J0054.9-7245,
XMMU~J005455.4-724512, CXOU~J005455.6-724510, SXP 504)
Small ROSAT error box  of this source contains
an emission-line star (809 in MA93) with typical Be star characteristics,
it is the brightest object in the area of localization.
A factor of five X-ray
flux variability strengthens the identification as Be/X-ray binary.
A probable binary period of 268 days has been detected in the
optical counterpart (\citealt{ecg2005b}). The relationship between
this orbital period and the pulse period of 504s is within
the normal variance found in the Corbet diagram (\citealt{c1984}).

\noindent
{\bf J0054.9-7226.} (2E 0053.2-7242, RX J0054.9-7226, 1WGA J0054.9-7226,
SAX J0054.9-7226, RX J0054.9-7227, XTE J0055-724)
RX J0054.9-7226 is known to be an X-ray binary pulsar with a pulse
period of $58.969 \pm 0.001$ s (\citealt{mlc98}; \citealt{scf98}).
\cite{lwc99} have suggested the orbital period equal to
65 days from subsequent
X-ray outbursts.
\cite{lcc04} have obtained the orbital period about
123 days based on the timing
analysis.
In the timing analysis of the XMM-Newton
data, the pulse period was verified to be $59.00 \pm 0.02$~s
(\citealt{sph03}). The
optical counterpart, a Be star, is identified with the variable star OGLE
J005456.17-722647.6
(\citealt{zsw01}).

\noindent
{\bf J005517.9-723853.} (XMMU J005517.9-723853, SXP 701)
This bright X-ray source was detected during XMM-Newton observation
of the SMC region around XTE J0055-727 (\citealt{hps04}).
The optical brightness and colors are consistent with  expectations
for a Be star companion, and the X-ray spectra are consistent with Be/X-ray binary.
Using MACHO and OGLE-II data, \cite{sc2004} obtained the data
showing a possible long-term period of 413 days, but further analysis is needed
to confirm it.

\noindent
{\bf J0055.4-7210.} (RX J0055.4-7210, 2E 0053.7-7227, CXOU
J005527.9-721058, WW 36)
Timing analysis on this object revealed a period of
$34.08 \pm 0.03$ s with a confidence of 98.5\%
(\citealt{ecg2004b}).
The position of this pulsar is within 3 arcsec of the
ROSAT source 2RXP J005527.1-721100. The latter is
coincident with a 16.8 V magnitude optical source
having a B-V color index of -0.116 (\citealt{zht2002})
which would be consistent with the value expected
from the optical companion in a Be/X-ray binary.

\noindent
{\bf 0054.4-7237.} (2E 0054.4-7237, XMMU J005605.2-722200, WW 38)
The error circle of the Einstein source 2E 0054.4-7237
cintains an emission line object. Therefore, it was suggested
as a Be/X-ray binary candidate (\citealt{sph03}).
In the XMM-Newton data, a source consistent with
the position of the emission line object was detected
(XMMU~J005605.2–722200) and pulsations from this source
were discovered (\citealt{sph03}). XMMU J005605.2-722200
is most likely consistent with 2E 0054.4-7237. The pulsar period
is $140.1 \pm 0.3$ s.

\noindent
{\bf J0057.4-7325.} (AX J0057.4-7325, RX J0057.4-7325)
Six ROSAT observations
have covered the position of AX J0057.4-7325.
Coherent pulsations with a barycentric period of $101.45
\pm 0.07$ s were discovered by \cite{ytk00} with ASCA. The flux
variability, the hard X-ray spectrum, and the long pulse period are
consistent with the hypothesis that AX~J0057.4-7325 is an X-ray binary
pulsar
with a companion which is either a Be, an OB supergiant, or a low-mass
star.
\cite{ytk00} found only one optical source, MACS J0057-734~10, in the
ASCA error circle. They note that OB supergiant
X-ray binaries in the SMC (only SMC X-1 and EXO 0114.6-7361) are both
located in the eastern wing and this fact may lead us to suspect that
AX J0057.4-7325 would be the third example.

\noindent
{\bf J005736.2-721934.} (CXOU J005736.2-721934, XMMU J005735.6-721934, XMMU
J005736.5-721936)
CXOU J005736.2-721934 was originally discovered
in Chandra observation in 2001 (\citealt{mfl03})
where it was reported to have a pulse period of 565.83s.
This X-ray source was also found by \cite{sph03}
in XMM-Newton EPIC data.
XMMU J005735.6-721934 has a hard spectrum
and positionally coincides with emission line
object 1020 in MA93.
This source is very faint during the XMM-Newton
observation and it is suggested as a new Be/X-ray candidate.

\noindent
{\bf J0057.8-7202.} (AX J0058-720, RX J0057.8-7202)
The pulse period of  AX J0058-720 was determined from the ASCA data as
$280.4 \pm 0.3$ s (\citealt{yk98b}).
 \cite{sph03} confirmed this value using the XMM-Newton data: 
$281.1 \pm 0.2$ s.
The source has been suggested to be a Be/X-ray candidate due to a likely
optical counterpart, which is an emission
line object.

\noindent
{\bf J0057.8-7207.} (CXOU J005750.3-720756, RX J0057.8-7207, XMMU
J005749.9-720756, XMMU J005750.3-720758)
This source is a Be/X-ray candidate with an emission
line object 1038 in MA93 suggested as a likely optical
counterpart (\citealt{hs00}).
\cite{sph03} discovered pulsations in the new XMM-Newton
data and derived a pulse period of $152.34 \pm 0.05$ s.
For this source, a pulsar period was
independently found in Chandra data by \cite{mfl03}.

\noindent
{\bf J0057.9-7156.} ( RX J0057.9-7156)
This source is a Be/X-ray binary candidate because of  a
positional coincidence 
with the emission-line object 1044 in MA93 (\citealt{hs00}).

\noindent
{\bf J0058.2-7231.} (RX J0058.2-7231, RX J0058.3-7229)
\cite{scc99} reported the detection of this very weak X-ray source by
ROSAT HRI. Its optical counterpart is a variable Be star in the SMC, OGLE
00581258-7230485 (\citealt{zsw01}).
\cite{scl2004b} have proposed the orbital period of 59.72 days using V, R
and I data from the MACHO and OGLE-II surveys.

\noindent
{\bf J0059.2-7138.}  (RX J0059.2-7138)
The supersoft source RX J0059.2-7138 was detected serendipitously with
the ROSAT PSPC in 1993 and was seen almost simultaneously by ASCA
(\citealt{hu94}; \citealt{ky96}). Previously, it had failed to be
detected by either the Einstein Observatory or EXOSAT in the early 1980s,
or in pointed ROSAT
observations of 1991. The transient nature of this source is clearly
established.
The best fit to the X-ray spectrum consists of three components
(\citealt{ky96}): two power laws with indices
0.7 and 2.0 fit the spectrum in the $> 3$ KeV and 0.5-3.0 keV bands
respectively. Furthermore, the emission is pulsed at levels of $ \sim 35\%$
and $ \sim 20 \%$ in these
respective bands, with a period of $ \sim 2.7$ s (\citealt{hu94}).
\cite{sc96} identified the probable optical counterpart of this source
with a 14th-magnitude B1~III emission star lying within the X-ray error
circle.

\noindent
{\bf J0059.3-7223.} (RX J0059.3-7223, XMMU J005921.0-722317)
This X-ray pulsar was discovered by \cite{mlm2004}.
There are two variable stars in both the OGLE (OGLE~151891)
and MCPS (MCPS 3345630) catalogs which are suggested as the
optical counterparts for this X-ray source. The angular
distance between these two catalog stars is only 0.3 arcseconds,
consistent with being the same source (\citealt{mlm2004}).
The absolute B magnitude (-4.1) of this star is approximately
consistent with a B0 star.

\noindent
{\bf J010030.2-722035.} (XMMU J010030.2-722035)
This X-ray source was found by \cite{sph03}
in XMM-Newton EPIC data.
XMMU~J010030.2-722035 has a hard spectrum
and positionally coincides with emission line
object 1208 in MA93.
This source was very faint during the XMM-Newton
observation.  It was suggested as a new Be/X-ray candidate.

\noindent
{\bf J0101.0-7206.} (RX J0101.0-7206, CXOU J010102.7-720658,
 XMMU J010103.1-720702,  XMMU J010102.5-720659)
The X-ray transient RX J0101.0-7206  was discovered in the course of
ROSAT observations of the SMC in October 1990 (\citealt{kp96}) at a
luminosity of $1.3\cdot 10^{36}\
\mathrm{erg~s^{-1}}$.
The source showed a luminosity of  $3\cdot 10^{33}\ \mathrm{erg s^{-1}}$
in the ROSAT band (0.1-2.4 keV) during two XMM-Newton observations
(\citealt{sph03}).
Pulsations with a period of $304.49  \pm 0.13$ s were discovered in Chandra
data
(\citealt{mfl03}). This period could not be verified in the XMM-Newton
observation, because the source was
too faint. \cite{ec03} presented results on the optical analysis of
likely counterparts, discussing two objects
(Nos. 1 and 4) in the ROSAT PSPC error circle. They conclude that the
optical counterpart is object No. 1 which is confirmed to be a Be star.

\noindent
{\bf J0101.3-7211.} (RX J0101.3-7211)
The source was detected in ROSAT observations and proposed by \cite{hs00}
as a Be/X-ray candidate.
The optical counterpart (OGLE 01012064-7211187) is a Be star.

\noindent
{\bf J0101.6-7204.} (RX J0101.6-7204)
\cite{hs00} suggested the identification of RXJ0101.6–7204
with object 1277 in MA93 from two accurate positions from
ROSAT HRI and PSPC observations. The factor of three
variability supports a Be/X-ray binary nature of this source.

\noindent
{\bf J0101.8-7223.} (AX J0101.8-7223,  XMMU J010152.4-722336)
\cite{hs00} suggested this source as a Be/X-ray binary.
They proposed the emission-line star 1288 in MA93 as a probable
optical counterpart. This star exhibits magnitudes
typical for a Be star in the SMC and is located near the
overlapping area of HRI and PSPC error circles.

\noindent
{\bf J0103-728.} (XTE J0103-728)
This source was detected with the RXTE Proportional
Counter Array (\citealt{cmm2003}).

\noindent
{\bf J0103-722.} (AX J0103-722,   2E 0101.5-7225,   SAX J0103.2-7209,
CXOU J010314.1-720915, 1E 0101.5-7226)
For the Be/X-ray binary  AX J0103-722 a pulse period of $345.2 \pm 0.1$~s
was determined by \cite{isc98}. In the XMM-Newton data, pulsations were
confirmed with a period of $341.7
\pm 0.4$ s (\citealt{sph03}). This source was detected with a nearly
constant flux  in all the
Einstein, ROSAT and ASCA pointings which surveyed the relevant region of
the SMC.

\noindent
{\bf J0103.6-7201.} (RX J0103.6-7201)
\cite{hs00} identified this source with object 1393
in MA93. RX J0103.6-7201 shows variability by a factor of
three between the ROSAT observations, consistent with
a Be/X-ray binary. Recently \cite{hp05} reported the discovery of 1323~s
periodicity of this source.

\noindent
{\bf J0104.1-7244.} (RX J0104.1-7244)
The most likely identification with emission-line
star 1440 in MA93 suggests RX J0104.1-7244 as a
Be/X-ray binary (\citealt{hs00}).

\noindent
{\bf J0104.5-7221.} (RX J0104.5-7221,  RX J0105.5-7221)
\cite{hs00} reported that this source was not detected by
the ROSAT PSPC but the accurate HRI position included only
the emission-line object 1470 from MA93 as a bright object
in the error circle.  RX J0104.5-7221 is therefore very
likely a Be/X-ray binary.

\noindent
{\bf J0105-722.} (AX J0105-722,  RX J0105.3-7210,   RX J0105.1-7211)
\cite{yk98c}  reported AX J0105-722 as an X-ray pulsar  with a
period of 3.34 s. From ROSAT PSPC images \cite{fhp00} resolved this source
into several
X-ray sources. They combined X-ray, radio-continuum and optical data to
identify the sources: for RX
J0105.1-7211 they proposed an emission line star from the catalogue of
Meyssonier \& Azzopardi in the X-ray error circle as a likely optical
counterpart.
This catalogue contains several known Be/X-ray binaries strongly
suggesting
RX J0105.1-7211 as a new Be/X-ray binary in the SMC.

\noindent
{\bf J0105.9-7203.} (RX J0105.9-7203, AX J0105.8-7203)
A single bright object (the emission-line star 1557 in MA93) 
was found in the small ROSAT PSPC error circle (source 120) (\citealt{hs00}), 
which made the identification of RX J0105.9–-7203 as Be/X-ray very likely.

\noindent
{\bf J0106.2-7205.} (SNR 0104-72.3, RX J0106.2-7205, 2E 0104.5-7221)
SNR 0104-72.3 contains a pointlike X-ray source with a blue optical
counterpart and $H_\alpha$ emission.

\noindent
{\bf J0107.1-7235.} (RX J0107.1-7235, AX J0107.2-7234,2E 0105.7-7251)
\cite{hs00} have identified this source with the emission-line
star 1619 in MA93. A Be/X-ray binary nature is likely.

\noindent
{\bf 0107-750.} (1H 0103-762,  H 0107-750)
This source is a very bright UV object with prominent $H_\alpha$ and
$H_\beta$ emission.

\noindent
{\bf J0111.2-7317.} (XTE J0111.2-7317, XTE J0111-732(?))
The X-ray transient  XTE J0111.2-7317 was discovered by the RXTE X-ray
observatory
in November 1998 (\citealt{clc98a}). Analysis of ASCA observation
(\citealt{cto98b}, \citealt{yit00}) identified this source as a 31 s
X-ray pulsar with a flux in the 0.7-10~keV band of $3.6\cdot 10^{-10}\,
\mathrm{erg~cm^{-2} s^{-1}}$ and $\sim$ 45\% pulsed fraction. The detection
was also confirmed from
the BATSE telescope on the CGRO satellite which detected the source in
the hard 20-50 keV band with a flux ranging from 18 to 30 mCrab
\citep{wf98}.
The source was not detected by ROSAT.
In the X-ray error box of   XTE J0111.2-7317 \cite{cnc01} found
a relatively bright object (V=15.4) which has been classified as a
B0.5-B1Ve star and that was later confirmed by \cite{chr00} as the most
plausible counterpart for
XTE J0111.2-7317. There is also evidence for the presence of a
surrounding
nebula, possibly a supernova remnant (\citealt{cnc01}).

\noindent
{\bf J0117.6-7330.} (RX J0117.6-7330)
This X-ray transient was discovered by the PSPC on
board ROSAT (\citealt{crw96}; \citealt{crw97}).
\cite{s99} conducted spectroscopic and photometric observations of the
optical companion of the
X-ray transient RX J0117.6-7330 during a quiescent state. The primary
component was identified as a B0.5~IIIe star.
\cite{mfh99} reported on the detection of pulsed, broadband, X-ray
emission from this transient source.
The pulse period of 22 s was detected by the ROSAT/PSPC instrument and by
the Compton Gamma-Ray Observatory/BATSE instrument. The total directly
measured X-ray
luminosity during the ROSAT observation
was $1.0\cdot 10^{38}\ \mathrm{erg~s^{-1}}$. The pulse frequency
increased
rapidly during the outburst with a peak spin-up
rate of $1.2\cdot 10^{-10}\ \mathrm{Hz~s^{-1}}$ and a total frequency
change of 1.8\%. The pulsed percentage
was 11.3\% from 0.1-2.5 keV, increasing to at least 78\% in the 20-70 keV
band. These results established RX J0117.6-7330 as a transient Be binary
system.

\noindent
{\bf J0119.6-7330.} (RX J0119.6-7330)
This source was detected once in the 0.9-2.0 keV band of the ROSAT PSPC.
An emission-line object in the error circle suggests an Be/X-ray binary
(\citealt{hs00}).

\noindent
{\bf J0119-731.} (XTE J0119-731)
This source was detected in the RXTE Proportional Counter
Array observations with intensity about 0.625 mCrab, and a
period of $2.1652 \pm 0.0001$ s (\citealt{cmm2003c}).
\cite{cg03} identified two emission-line optical counterparts
were first identified by searching the XTE error box using
SIMBAD : 1864 in MA93 and Lin 526. The second source, Lin 526,
exhibited strong H alpha and H beta emission. \cite{cg03} proposed
Lin 526 as the most likely counterpart to XTE J0119-731.

\noindent
{\bf SXP 46.4.} (SXP 46.4)
This source was detected in the RXTE Proportional
Counter Array observations.
The source position is not accurately known.

\noindent
{\bf SXP 89.} (SXP 89)
This source was detected in the RXTE Proportional
Counter Array observations.
The source position is not accurately known.

\noindent
{\bf XTE SMC144s.} (XTE SMC144s)
The source position is not accurately known.
\cite{clm2003} have detected this transient X-ray pulsar in
the Small Magellanic Cloud with the RXTE Proportional Counter
Array. They interpreted the outburst recurrence period as
the orbital period of a neutron/Be star binary with outbursts
occurring at periastron passage.

\noindent
{\bf SXP 165.} (SXP 165)
This source was detected in the RXTE Proportional
Counter Array observations.
The source position is not accurately known.


\section{Graphs and discussion}

In this section we present several useful plots based on the data from the
tables above.
In the first figure we show a usual period -- luminosity dependence.
If luminosity is proportional to $\dot M$~--~an accretion rate~--~ then
for each value of $L$ it is possible to determine a critical period,
$P_\mathrm{A}$ (see details on the magneto-rotational evolution of neutron
stars, for example, in \citealt{l92}).
It is determined by an equality of the magnetospheric radius
to the corotation radius, so $P_\mathrm{A}$ depends also on the magnetic field
of a neutron star. If a spin period of a neutron star is shorter than $P_\mathrm{A}$
then accretion rate is significantly reduced, and the neutron star is at the stage of
{\it propeller}. Lines for $P_\mathrm{A}$ for two values of the magnetic field
are shown in the figure. Situation can be more complicated for low accretion
rates when the so-called {\it subsonic propeller} stage becomes important.
In that case for a neutron star it is necessary  to slow down to a new
critical period $P_\mathrm{crit}$. Lines for this quantity are also shown
(see figure caption for other details).

In the second figure we present the so-called "Corbet diagram" (\citealt{c86}).
For most of Be/X-ray binaries the
correlation between spin and orbital periods is
strong, so that this dependence is even used to estimate orbital periods
when only spins are known.



The observational number distributions of Be/X-ray binaries over orbital
characteristics are shown in figs. 3 and 4. It is clear that Be-systems do
not have orbital periods longer than one year.
There is a lack of systems with periods
10--20 days. As it was shown in the
paper by \cite{rl1998} the lack of short-period
Be/X-ray binaries can be explained by the effect of tidal synchronization in
binaries. The peak of the observed number distribution of Be/X-ray
systems over eccentricities falls in the range $0.4-0.5$.
In order to get a better agreement with the observed
parameters of Be/X-ray binaries there is no necessity of high kicks.
Moderate recoil velocities of the order  50 km s$^{-1}$ are enough
(see \citealt{rl1998}). It can be a particular feature of Be/X-ray binaries
\citep{petal2004}.

\begin{figure}
\includegraphics[width=174pt]{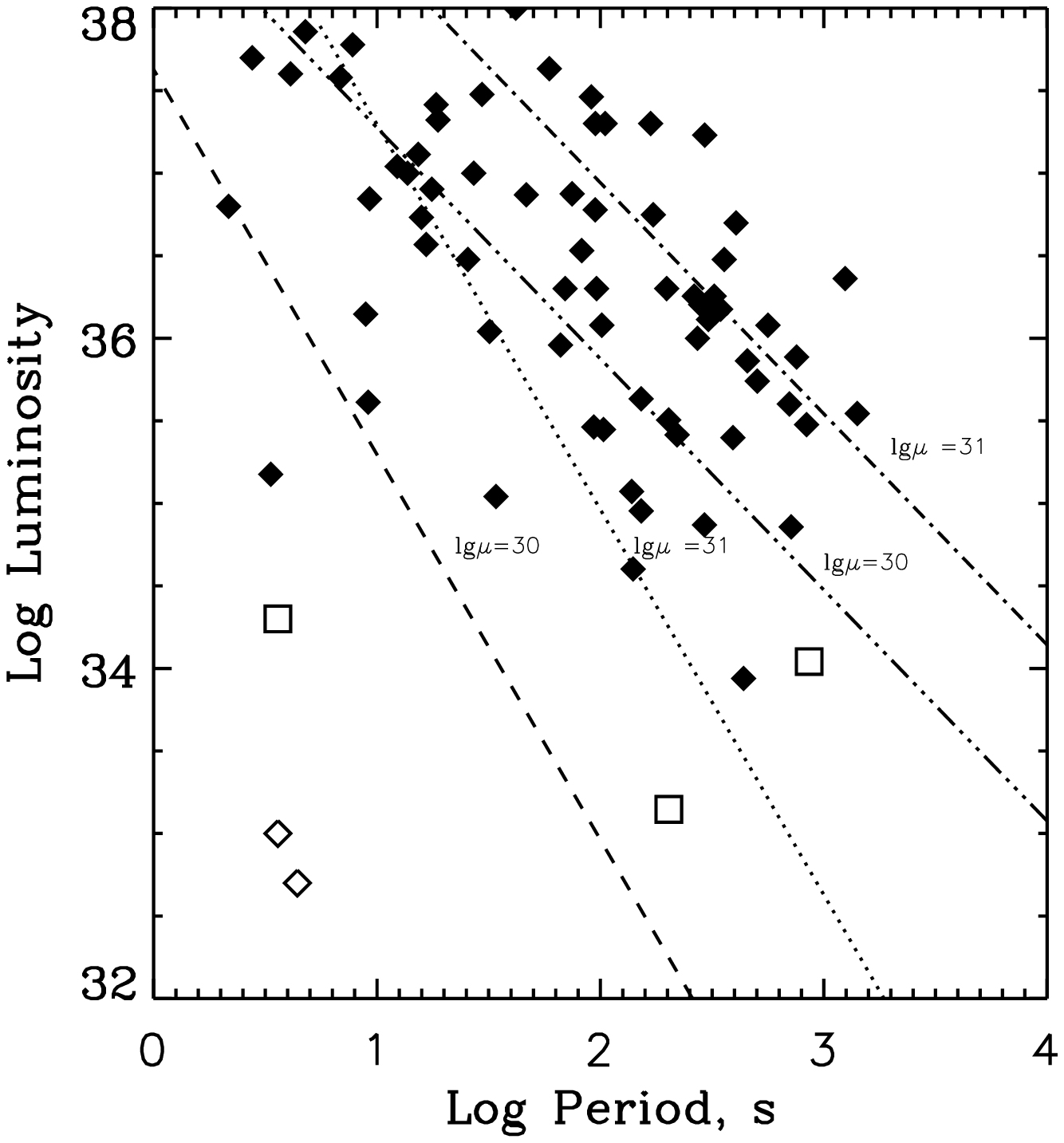}
  \caption{Period~--~Luminosity.
Open symbols correspond to the quiescent state
of the X-ray pulsar.
Squares represent three sources in quiescence from
which pulsations were observed (4U 0115+63 -- \citet{c2001};
RX J0440.9+4431 and RX J1037.5-564 -- \citet{rr99}).
Open diamonds show objects without pulsations in quiescence,
which are supposed to be
in the {\it propeller} state (4U 0115+63 and V0332+53 -- \citet{csi02}).
The graph is artificially truncated at log~$p=0$ and
log~$L=38$.
So, here we do not plot
two systems with the most fastly rotating neutron stars: 
J0635+0533 (small luminosity)
and 0535-668 (large luminosity).
Dashed and dotted lines correspond to the
critical period,
$P_\mathrm{A}=2^{5/14}\pi (GM)^{-5/7} (\mu^2/\dot M)^{3/7}$, for two values
of the magnetic moment, $\mu=10^{30}$~G~cm$^3$
and $10^{31}$~G~cm$^3$. The
two dashed-dotted lines correspond to subsonic propeller~--~accretor
transition for the same two values of the magnetic moment which
occurs at
$P_\mathrm{crit}=81.5 \mu_{30}^{16/21} L_{36}^{-5/7}$ according to 
\citet{i2003}.
We note that the multiplicative coefficient in Ikhsanov's formula is
larger than in the classical formula of
\citet{dp1981} by a factor $\sim7.5$.
}
\label{fig:pl}
\end{figure}

\begin{figure}
\includegraphics[width=174pt]{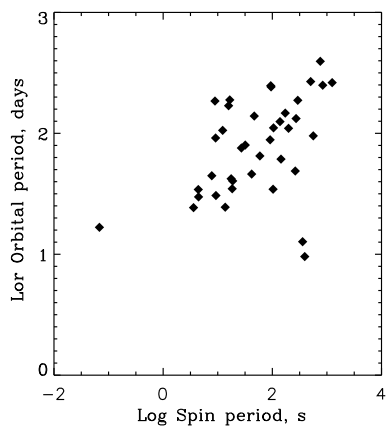}
  \caption{Spin period~--~Orbital period. Data points for 39 systems are shown. 
Three displaced systems are:
2206+543 (large spin and short orbital periods),
2103.5+4545 (large spin and short orbital periods) and 0535-668 (very short
spin period).
}
\label{fig:pporb}
\end{figure}




\begin{figure}
\includegraphics[width=174pt]{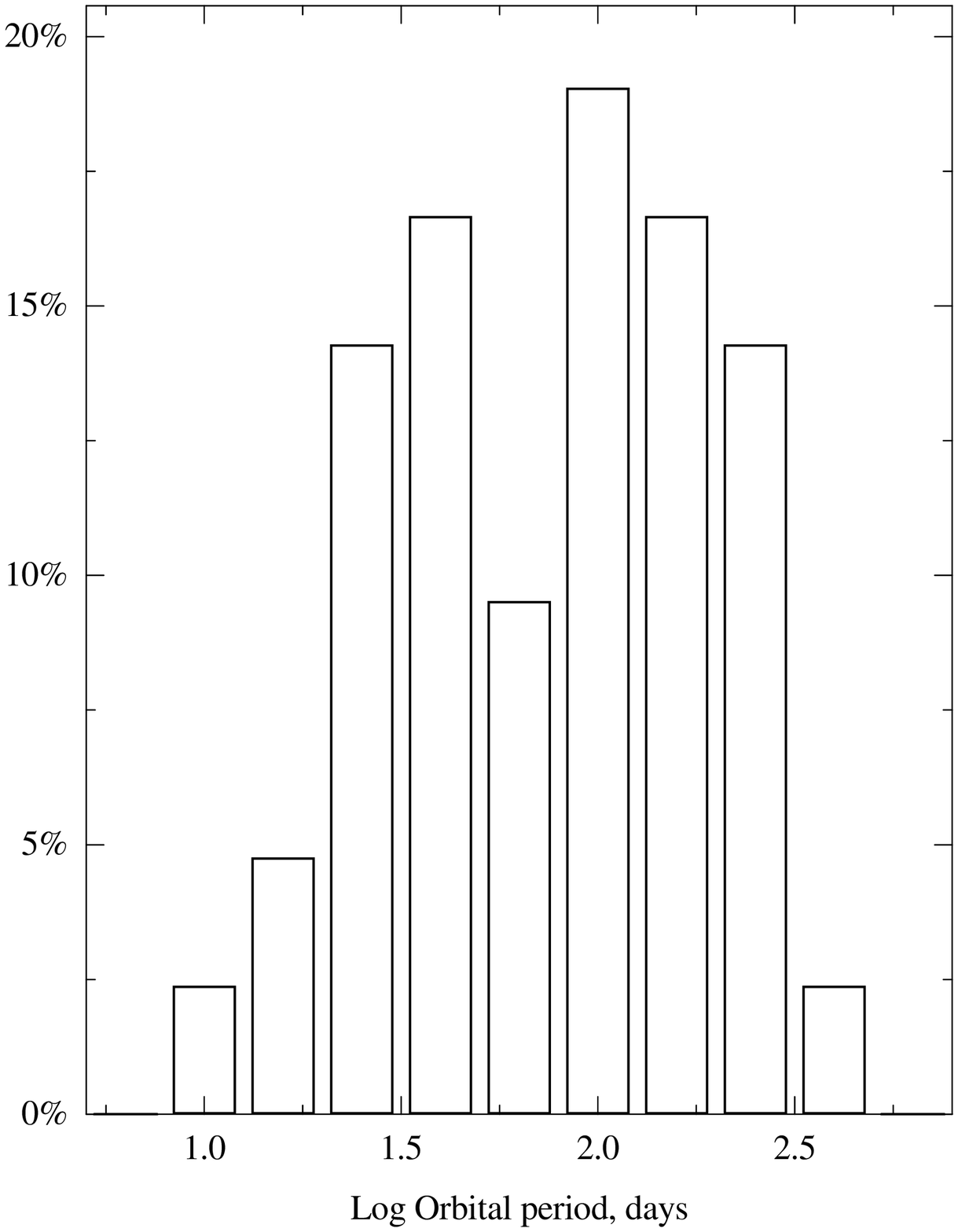}
  \caption{The observational number distribution of Be/X-ray binaries over
orbital period.}
\label{fig:gist_porb}
\end{figure}

\begin{figure}
\includegraphics[width=174pt]{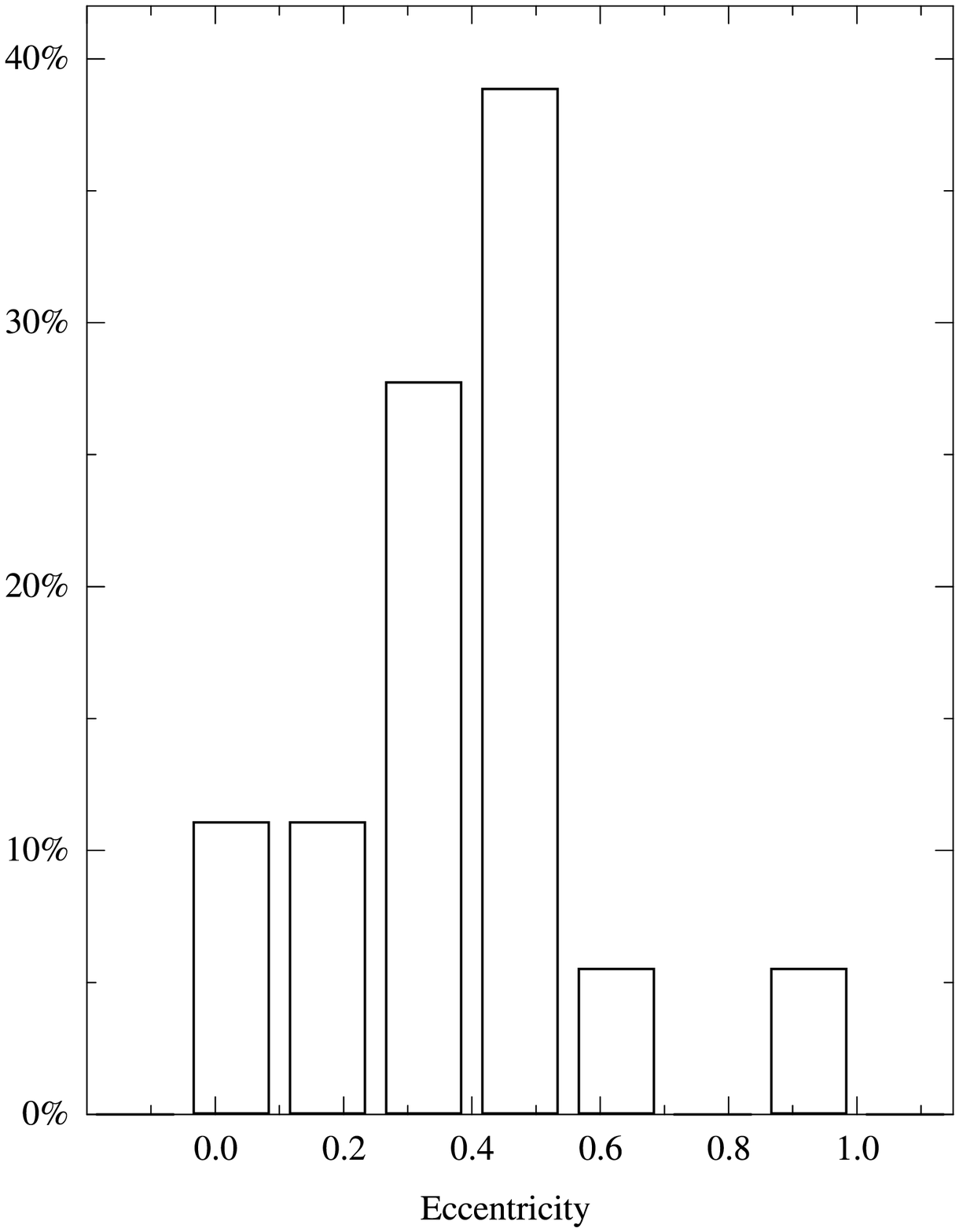}
  \caption{The observational number distribution of Be/X-ray binaries over
orbital eccentricity}
\label{fig:gist_ecc}
\end{figure}



\section*{Acknowledgments}

The work was supported by the
Russian Foundation for Basic Research (RFBR)
grants 03-02-16068 and 04-02-16720.
S.P. thanks the ``Dynasty'' Foundation (Russia).



\bsp

\label{lastpage}

\end{document}